\documentstyle[12pt]{article}

\begin{document}
\hfill\vbox{
\hbox{OHSTPY-HEP-T-97-012}
\hbox{cond-mat/9707199}
\hbox{July 18 1997} }\par
\thispagestyle{empty}

\vspace{.5in}

\begin{center}
{\Large \bf Quantum Corrections to the Ground State 
	of a Trapped Bose-Einstein Condensate}

\vspace{.5in}

Eric Braaten and Agustin Nieto\footnote{ 
	Address after September 1, 1997:
	CERN -- Theory Division, CH-1211 Geneva 23,
	Switzerland.}
 \\

{\it Department of Physics, The Ohio State University, Columbus, OH 43210}

\end{center}

\vspace{.5in}

\begin{abstract}
In the mean-field approximation,
the number density $\rho({\bf r})$ for 
the ground state of a Bose-Einstein condensate
trapped by an external potential $V({\bf r})$ satisfies a 
classical field equation called the Gross-Pitaevskii equation.
We show that quantum corrections to $\rho$ are dominated by 
quantum fluctuations with wavelengths of order 
$1/\sqrt{\rho a}$, where $a$ is the S-wave scattering length.
By expanding the equations for the Hartree-Fock approximation 
to second order in the gradient expansion, we derive
local correction terms to the Gross-Pitaevskii equation
that take into account the dominant effects of quantum fluctuations.  
We also show that the gradient expansion for the density 
breaks down at fourth order.
\end{abstract}

\newpage

\section{Introduction}
\label{sec:Intro}

The successful achievement of Bose-Einstein condensation of atomic
gases in magnetic traps \cite{Colorado,Rice,MIT} 
has created an explosion of interest in interacting Bose gases.
The condensates in existing magnetic traps are sufficiently dilute
that the mean-field approximation gives a satisfactory description of 
present experimental measurements.
However accurate theoretical predictions require that 
quantum fluctuations around the mean field also be taken into account.
The relative magnitude of these corrections grows as the 
square root of the number density of the atoms.
They will therefore become more important as higher
trap densities are achieved and as the precision of 
experimental measurements improves.

One of the basic observables of a Bose-Einstein condensate
trapped in an external potential $V({\bf r})$ 
is the number-density profile $\rho({\bf r})$ of the ground state.
In the mean-field approximation, $\rho({\bf r})$ satisfies
the Gross-Pitaevskii equation:
\begin{equation}
\left(  {\hbar^2 \over 2 m} \mbox{\boldmath $\nabla$}^2 
	+ \mu - V({\bf r}) \right) \sqrt{\rho}({\bf r})
	\;-\; {4 \pi \hbar^2 a \over m} \rho \sqrt{\rho}({\bf r})
\;=\; 0 \,,
\label{GP-rho}
\end{equation}
where $a$ is the S-wave scattering length of the atoms.  The chemical potential 
$\mu$ must be tuned so that $\int d^3 r \, \rho = N$, 
where $N$ is the number of atoms in the trap.
The density profile of a trapped Bose-Einstein condensate
has been studied extensively using (\ref{GP-rho}).
The solutions to this equation have been calculated using 
numerical methods \cite{Edwards-Burnett,Dalfovo-Stringari}
and variational methods \cite{Baym-Pethick,Dodd,Ilinski-Moroz,Fetter}.
The solutions have also been studied analytically in the Thomas-Fermi limit,
in which the gradient term in (\ref{GP-rho}) is neglected
\cite{Baym-Pethick}.  The corrections due to the breakdown of this
approximation near the edge of the condensate have also been studied 
\cite{D-P-S,Fetter-Feder}. 

There are corrections to the mean-field approximation 
from quantum fluctuations 
around the mean field.  
In a dilute homogeneous Bose gas,
the relative magnitude of the contributions of quantum fluctuations 
to thermodynamic quantities
is characterized by the dimensionless quantity $\sqrt{\rho a^3}$.
For condensates in existing magnetic traps, 
the peak value of  $\sqrt{\rho a^3}$ is small 
but not negligible.  
Since there are some observables that are 
more sensitive than the density to the effects of quantum fluctuations,
it is important to be able to calculate the effects of 
quantum fluctuations
quantitatively.
 
In this paper, we calculate the effects of quantum fluctuations 
on the density profile 
for a Bose-Einstein condensate in a trapping potential.
The expansion parameter that characterizes the relative
magnitude of these effects is  
$\sqrt{\rho a^3}$,  where $\rho$ is the local number density.
We point out that the quantum corrections are dominated by 
quantum fluctuations with wavelengths on the order of $1/\sqrt{\rho a}$.
The leading effects of these short-distance quantum fluctuations
can be calculated using the gradient expansion.
By carrying out a self-consistent one-loop calculation
through second order in the gradient expansion,
we determine the correction terms that must be added to the 
Gross-Pitaevskii equation (\ref{GP-rho}) to take into account 
the effects of quantum fluctuations:
\begin{eqnarray}
0 &=& 
\left( {\hbar^2 \over 2 m} \mbox{\boldmath $\nabla$}^2 
	+ \mu - V({\bf r}) \right) \sqrt{\rho}({\bf r})
\;-\; {4 \pi \hbar^2 a \over m} \rho \sqrt{\rho}({\bf r})
\nonumber \\
&&  
\;-\; {128 \sqrt{\pi} \hbar^2 a^{5/2} \over 3 m} \rho^2({\bf r})
\;-\; {17 \hbar^2 a^{3/2} \over 18 \sqrt{\pi} m} 
\left[ 2 \sqrt{\rho} \, \mbox{\boldmath $\nabla$}^2 \sqrt{\rho} ({\bf r})
\;+\; ( \mbox{\boldmath $\nabla$} \sqrt{\rho} )^2({\bf r}) \right]\,.
\label{rho-final}
\end{eqnarray}
Our method involves a combination of the Hartree-Fock approach \cite{Griffin}
and the Thomas-Fermi approach \cite{C-Y-Y,T-T-H}.  
In the Hartree-Fock approximation, which involves 
the self-consistent treatment of one-loop quantum corrections,
the equation for the density is an integral equation.
We obtain the local differential equation (\ref{rho-final})
by applying a gradient expansion to the integral equation,
which corresponds to expanding around the Thomas-Fermi limit.

We begin in Section \ref{sec:QFT} by formulating the problem 
of Bose-Einstein condensation in a trapping potential as a problem 
in quantum field theory.   
In Section \ref{sec:Cartesian},
we develop a perturbation expansion for calculating the effects of 
quantum fluctuations around an arbitrary background field. 
In Section \ref{sec:One-loop},
we calculate the one-loop quantum corrections to the density
profile and the condensate profile.
We show that the ultraviolet divergences 
that arise in the calculation can be removed by 
the same renormalizations of the action and the number density
that are required in the absence of the potential.
We find that the number density can be expanded 
to second order in the gradient of the mean field,
while the gradient expansion for the condensate breaks down at that order.
In Section \ref{sec:Self-consistent}, we calculate the 
self-consistent one-loop quantum corrections to the density profile
to second order in the gradient expansion 
and show that they are given by (\ref{rho-final}).
We repeat the calculation in Section \ref{sec:Polar} using an 
alternative parameterization for the quantum field and demonstrate 
that the final result is independent of the parameterization. 
We also use this parameterization to show that the gradient expansion 
for the density breaks down at fourth order.
Finally, in Section \ref{sec:Imp}, we examine the 
implications of our results for Bose-Einstein condensation in
present magnetic traps.  Details of the calculations of Feynman diagrams
are included in several appendices.

\section{Quantum field theory formulation}
\label{sec:QFT}

Consider a large number $N$ of identical bosonic atoms trapped in an 
external potential $V({\bf r})$ at zero temperature.
If the  momenta of the atoms are sufficiently low, their de Broglie 
wavelengths are much smaller than the range of the interatomic 
potential, which is comparable in magnitude to the Bohr radius.  
In this case, the scattering of two atoms
will be dominated by S-wave scattering and can be described by a single
number, the S-wave scattering length $a$.
Our problem is to determine the number-density profile 
$\rho({\bf r})$ of the atoms.
We begin by formulating this many-body quantum mechanics problem 
as a problem in quantum field theory.

A convenient way to describe a system containing any number $N$ of atoms
is in terms of a quantum field $\psi({\bf r},t)$ that annihilates an atom.
If the atom is a boson, the field satisfies the equal-time 
commutation relations 
\begin{eqnarray}
\Big[ \psi({\bf r},t), \psi({\bf r}',t) \Big] 
&=& 0 \,,
\label{ccr-psi}
\\
\left[ \psi({\bf r},t), \psi^\dagger({\bf r}',t) \right] 
&=& \delta({\bf r}-{\bf r}') \,.
\label{ccr-psidag}
\end{eqnarray}
The time evolution of the field is given by the equation
\begin{equation}
i \hbar {\partial \ \over \partial t} \psi
\;=\; \left[ - {\hbar^2  \over 2 m} \mbox{\boldmath $\nabla$}^2 
	+ V({\bf r}) \right]  \psi 
\;+\;  {g + \delta g \over 2} \psi^\dagger \psi \psi \,,
\label{NLSchr}
\end{equation}
where the coupling constant $g$ is related to the S-wave 
scattering length $a$ by 
\begin{equation}
g \;=\; {8 \pi \hbar^2 a \over m} \,.
\label{g-a}
\end{equation}
The parameter $\delta g$ in (\ref{NLSchr}) is a counterterm
associated with the renormalization of $g$.
It is necessary to impose an ultraviolet cutoff $\Lambda_{\rm UV}$
on the wavenumbers of virtual particles in order to avoid
ultraviolet divergences due to quantum fluctuations of the field.
Renormalization  of a quantum field theory is the tuning of its 
parameters so that physical quantities are independent of the 
ultraviolet cutoff.  All the dependence of first order quantum 
corrections on $\Lambda_{\rm UV}$ can be removed by adjusting  
$\delta g$ in (\ref{NLSchr}) as a function of $\Lambda_{\rm UV}$. 

The number operator, which counts the number of atoms, is
\begin{equation}
{\cal N} \;=\;  \int d^3r \, \psi^\dagger \psi({\bf r},t) \,.
\label{number}
\end{equation}
That this is a number operator follows from the commutation relations 
(\ref{ccr-psi}) and (\ref{ccr-psidag}), which
imply that $\psi^\dagger$ and $\psi$ act as raising and lowering 
operators for ${\cal N}$. 
The equation (\ref{NLSchr}) implies that ${\cal N}$
is independent of time, so the number of atoms is conserved. 
The equation (\ref{NLSchr}) can also be expressed in the form
\begin{equation}
i \hbar {\partial \ \over \partial t} \psi
\;=\; - \left[ {\cal H}, \psi \right] \,,
\label{NLSchr-H}
\end{equation}
where the hamiltonian operator ${\cal H}$ is 
\begin{equation}
{\cal H} \;=\; 
\int d^3r \, \left(
\psi^\dagger \left[ - {\hbar^2  \over 2 m} \mbox{\boldmath $\nabla$}^2 
	+ V({\bf r}) \right]  \psi 
\;+\;  {g + \delta g \over 4} \psi^\dagger \psi^\dagger \psi \psi 
\right) \,.
\label{H}
\end{equation}
The hamiltonian ${\cal H}$ measures the energy of the atoms.
  
The vacuum $| 0 \rangle$, defined by $\psi({\bf r},t) | 0 \rangle = 0$, 
represents the state containing zero atoms.
One can show that a  
Schroedinger wavefunction for $N$ atoms
can be represented as a matrix element of an
operator between a state with ${\cal N}=N$ and the vacuum.
The simplest case is
a state $| \phi \rangle$ containing 1 atom, which satisfies
${\cal N}| \phi \rangle = | \phi \rangle$.
Since the last term in (\ref{NLSchr}) annihilates the 
single particle state $| \phi \rangle$,
the matrix element $\langle 0 | \psi({\bf r},t) | \phi \rangle$
satisfies the Schroedinger equation: 
\begin{equation}
\left[ i \hbar {\partial \ \over \partial t}
	+ {\hbar^2  \over 2 m} \mbox{\boldmath $\nabla$}^2 
	- V({\bf r}) \right] 
\langle 0 | \psi({\bf r},t) | \phi \rangle
\;=\;  0 \,.
\label{Schr}
\end{equation}
Thus $\langle 0 | \psi({\bf r},t) | \phi \rangle$
can be interpreted as the Schroedinger wavefunction for 
an atom in the potential $V({\bf r})$.

The next simplest case is 
a state $| \phi \rangle$ containing 2 atoms, which satisfies
${\cal N}| \phi \rangle = 2 | \phi \rangle$.
It is straightforward to show using (\ref{NLSchr}) that
the matrix element 
$\langle 0 | \psi({\bf r}_1,t) \psi({\bf r}_2,t) | \phi \rangle$
satisfies the Schroedinger equation for two particles
in the external potential $V({\bf r})$ interacting through 
a two-body potential proportional to $\delta^3({\bf r}_1-{\bf r}_2)$.
In the absence of the potential $V({\bf r})$, 
one can calculate the amplitude for the scattering of two atoms
exactly \cite{Jackiw}.  The scattering amplitude $f(\theta)$ 
is independent of the scattering angle $\theta$, 
so it describes S-wave scattering.
If the total energy of the two atoms in the 
center-of-momentum frame is $E=2(p^2/2m)$, 
the scattering amplitude is
\begin{equation}
f \;=\;  
- {1 \over 4 \pi} \left[ {2 \hbar^2 \over m ( g + \delta g)} 
\;+\; \int {d^3k \over (2 \pi)^3} {1 \over k^2 - mE/\hbar^2 - i \epsilon}
	\right]^{-1} \,.
\label{f-int}
\end{equation}
The integral over the wave-vector ${\bf k}$ 
is ultraviolet divergent.  A particularly convenient regularization
of the integral is to introduce an ultraviolet cutoff $\Lambda_{\rm UV}$
through the following prescription:
\begin{eqnarray}
\int {d^3k \over (2 \pi)^3} {1 \over k^2 - mE/\hbar^2 - i \epsilon}
&\equiv& {1 \over 2 \pi^2} \int_0^\infty k^2 dk
	\left( {1 \over k^2 - mE/\hbar^2 - i \epsilon} - {1 \over k^2} \right)
\nonumber \\
&&
\;+\; {1 \over 2 \pi^2} \int_0^{\Lambda_{\rm UV}} k^2 dk \; {1 \over k^2}
	\,,
\label{UV-reg}
\end{eqnarray}
The scattering amplitude then becomes
\begin{equation}
f \;=\;  
- {1 \over 4 \pi} \left[ {2 \hbar^2 \over m ( g + \delta g)} 
	\;+\; {1 \over 2 \pi^2} \Lambda_{\rm UV}
	\;+\;  i {\sqrt{m E} \over 4 \pi \hbar} \right]^{-1} \,.
\label{f-unren}
\end{equation}
The dependence on the ultraviolet cutoff can be completely cancelled 
by choosing the bare coupling constant $g + \delta g$ to be
\begin{equation}
g + \delta g \;=\;  
{g \over 1 \;-\; m g \Lambda_{\rm UV} / (2 \pi \hbar)^2 }\,.
\label{g-bare}
\end{equation}
The scattering amplitude (\ref{f-unren}) then reduces to
\begin{equation}
f \;=\; - {a \over 1 + i a \sqrt{mE}/\hbar} \,.
\label{f-ren}
\end{equation}
This confirms the identification of $a$ as the S-wave scattering length.   
The scattering of atoms is correctly reproduced by the
pointlike interaction term in (\ref{NLSchr})
as long as the energy of the atoms is sufficiently low that 
(\ref{f-unren}) is a good approximation to the scattering amplitude.
Note that the energy dependence in (\ref{f-ren}) is that required
by the optical theorem. 

It is sometimes stated that a delta-function potential in 3 dimensions
is trivial in the sense that it gives no scattering.  
A more accurate statement is that there is no 
scattering if we take the ultraviolet cutoff to infinity with 
the strength of the potential held fixed. 
This is evident from (\ref{f-unren}),
which shows that $f \to 0$ if we take $\Lambda_{\rm UV} \to \infty$
with $g + \delta g$ fixed.  However, if we allow the strength 
of the potential to vary with $\Lambda_{\rm UV}$ in accordance with
(\ref{g-bare}), we obtain the nontrivial scattering amplitude (\ref{f-ren}).

 From the expression (\ref{f-unren}) for the scattering amplitude,
one can infer an upper limit on the ultraviolet cutoff that must be satisfied
in order for perturbative calculations to be accurate.
The expansion for $f$ in powers of $g$, including
the first quantum correction which is proportional to $g^2$, is
\begin{equation}
f \;=\;  
- {m g \over 8 \pi \hbar^2}
\left[ 1 \;+\;
\left( {m g \Lambda_{\rm UV} \over (2 \pi \hbar)^2}  - \delta g
	+  i {m g \sqrt{m E} \over 8 \pi \hbar^3} \right)
\;+\; \ldots \right] \,.
\label{f-pert}
\end{equation}
If the ultraviolet cutoff $\Lambda_{\rm UV}$ is too large, 
there is a delicate cancellation between the term proportional to 
$\Lambda_{\rm UV}$ in (\ref{f-pert}), which comes from an integral over $k$, 
and the counterterm $\delta g$.  Since a perturbative calculation 
is necessarily approximate, the cancellation can lead to large errors.
Such a large cancellation can be avoided if the term proportional to 
$\Lambda_{\rm UV}$ in (\ref{f-pert}) is much less than 1. 
This sets an upper bound on the ultraviolet cutoff:
\begin{equation}
\Lambda_{\rm UV} \;\ll\; {(2 \pi \hbar)^2 \over m g} 
	\;=\; {\pi \over 2 a} \,.
\label{LambdaUV-max}
\end{equation}
If this upper bound on $\Lambda_{\rm UV}$ is not satisfied, 
then in order to obtain an accurate calculation, it is necessary to use a
nonperturbative calculational method that sums up all orders in $g$.

A state $| \phi \rangle$ containing $3$ atoms 
satisfies ${\cal N} | \phi \rangle = 3 | \phi \rangle$.
In the absence of the potential $V({\bf r})$,
one can calculate the amplitude for $3 \to 3$ scattering 
as an expansion in powers of $g$.  The leading contribution is 
proportional to $g^2$ and comes from two successive $2 \to 2$ scatterings.
Higher order terms in $g$ represent quantum corrections.
The dependence of the first quantum correction on the ultraviolet cutoff
$\Lambda_{\rm UV}$ is cancelled by the counterterm $\delta g$
in (\ref{NLSchr}). However, the second quantum correction, which 
is proportional to $g^4$ has a logarithmic ultraviolet divergence 
that is not cancelled \cite{Braaten-Nieto}.  
Thus corrections to physical quantities from second order in the 
quantum fluctuations depend on the ultraviolet cutoff.  
One can eliminate the dependence on $\Lambda_{\rm UV}$ from second order 
quantum corrections by adding to (\ref{NLSchr}) the term 
$(g_3 + \delta g_3)  \psi^\dagger \psi^\dagger \psi \psi \psi/18$.
The logarithmic ultraviolet divergence is cancelled by
choosing the counterterm to be 
\begin{equation}
\delta g_3
\;=\; {3 (4 \pi - 3 \sqrt{3}) \over 32 \pi^3} m^3 g^4 
	\log {\Lambda_{\rm UV} \over \mu} \,,
\label{delg3}
\end{equation}
where the value of $\mu$, which was introduced to make the argument 
of the logarithm dimensionless, depends on the precise definition of $g_3$.
The parameter $g_3$ represents a point-like contribution to the 
$3 \to 3$ scattering amplitude. 
Thus the S-wave scattering length $a$ does not 
contain enough information about the low-energy scattering of atoms to
calculate second order quantum fluctuations. 
It is also necessary to know the $3 \to 3$ coupling constant $g_3$.
In this paper, we will avoid this complication by calculating only to 
first order in the quantum corrections.

A state $| \phi \rangle$ containing $N$ atoms 
satisfies ${\cal N} | \phi \rangle = N | \phi \rangle$.  
In the presence of the potential $V({\bf r})$,
the ground state in the ${\cal N} = N$ sector,
which we denote by $| \Omega_N \rangle$, 
is the state that minimizes $\langle \phi | {\cal H} | \phi \rangle$
subject to the constraint ${\cal N} | \phi \rangle = N | \phi \rangle$.
The desired number-density profile is  
\begin{equation}
\rho({\bf r}) \;=\; 
\langle \Omega_N | \psi^\dagger \psi({\bf r}) | \Omega_N \rangle\,.
\label{rho-def}
\end{equation}
If $N$ is large, we expect $\rho({\bf r})$ to be insensitive to 
changes in $N$ that are small compared to $N$. 
This suggests that we can relax the constraint on the particle number
and replace the state $| \Omega_N \rangle$ in (\ref{rho-def})
by the state that minimizes $\langle \phi | {\cal H} | \phi \rangle$
subject to the weaker constraint $\langle \phi | {\cal N} | \phi \rangle = N$.
If the root-mean-square fluctuations of ${\cal N}$ 
in that state are small compared to $N$, the 
expectation value of $\psi^\dagger \psi({\bf r})$ in that state 
should give a good approximation to (\ref{rho-def}).
But that state is precisely the ground state $| \Omega_\mu \rangle$
of the quantum field theory whose hamiltonian is ${\cal H} - \mu {\cal N}$,
where the value of the chemical potential $\mu$ is such that 
\begin{equation}
\langle \Omega_\mu | {\cal N} | \Omega_\mu \rangle
\;=\;  N \,.
\label{mu-N}
\end{equation}
Thus, if $N$ is sufficiently large,
the density profile can be approximated by the ground-state
expectation value of the operator $\psi^\dagger \psi({\bf r})$ 
in the state $| \Omega_\mu \rangle$.

We have now formulated the problem of calculating the density 
profile as a quantum field theory problem.  The field theory is
summarized by the action
\begin{equation}
S[\psi] \;=\;  \int dt \int d^3r \left\{  
\psi^\dagger \left[ i \hbar {\partial \ \over \partial t}
	+ {\hbar^2  \over 2 m} \mbox{\boldmath $\nabla$}^2 
	+ (\mu + \delta \mu) - V({\bf r}) \right] \psi  
\;-\;  {g + \delta g \over 4} \left(\psi^\dagger \psi \right)^2 \right\} \,.
\label{action}
\end{equation}
The counterterms $\delta \mu$ and $\delta g$ are needed to cancel ultraviolet 
divergences associated with quantum fluctuations of the field.
The counterterm $\delta \mu$ would also have been required in 
(\ref{NLSchr}) if the $\psi^\dagger \psi \psi$ term had not been 
normal-ordered.  The local number-density operator
$\psi^\dagger \psi({\bf r})$ can also have ultraviolet divergences,
and a counterterm $\delta \rho$ is therefore needed to cancel 
these divergences.  The number-density profile is the expectation value 
of this local operator in the ground state of the field theory: 
\begin{equation}
\rho({\bf r}) \;=\; 
\langle \psi^\dagger \psi({\bf r}) \rangle  \;-\; \delta \rho \,.
\label{rho-QFT}
\end{equation}
Here and below, we use the angular brackets $ \langle \ldots \rangle$ 
to denote the expectation value in the ground state $| \Omega_\mu \rangle$.
The chemical potential $\mu$ in (\ref{action})
must be adjusted so that the integral of the local number density 
is equal to the number of atoms:
\begin{equation}
\int d^3r \; \rho({\bf r}) \;=\; N \,.
\label{mu-def}
\end{equation}
Thus our problem reduces to calculating the ground-state expectation 
value (\ref{rho-QFT}) for the quantum field theory defined by (\ref{action}).
Another important observable is the condensate 
$\langle \psi({\bf r}) \rangle$, which is the ground-state expectation 
value of the field.  A nonzero value of the condensate indicates 
the spontaneous breaking of the phase symmetry $\psi \to e^{i \alpha} \psi$
of the action (\ref{action}).

The ultraviolet divergences that are cancelled by the counterterms 
$\delta \mu$, $\delta g$, and $\delta \rho$ arise from  treating
the interaction between atoms as pointlike down to arbitrarily 
short distances.  The divergences could be avoided by replacing
the pointlike $(\psi^\dagger \psi)^2$ interaction term in the 
action (\ref{action}) by an interaction through a two-body potential
$v({\bf r}_1 - {\bf r}_2)$.  A physically realistic  
two-body potential would have a range comparable to the size 
of an atom and its shape would have to be adjusted so that it 
gives the correct S-wave scattering length $a$.
It would be rather inefficient to calculate the effects of interactions 
using a physically realistic two-body potential. 
The reason is that physical quantities depend on the two-body
potential in a very simple way.  Almost all of the dependence enters 
through the S-wave scattering length $a$.
Thus we can obtain the same result for physical quantities
by using  a pointlike interaction that gives the same scattering length.
A pointlike interaction necessarily leads to ultraviolet divergences from
short-distance quantum fluctuations,
and they must be cancelled by counterterms.
Since the counterterms cancel short-distance effects, they correspond to
renormalization of the parameters in the action and to renormalization
of local operators, such as $\psi^\dagger \psi({\bf r})$.
We will find that the three counterterms
$\delta \mu$, $\delta g$, and $\delta \rho$ are sufficient to
cancel all the ultraviolet divergences associated with first order 
quantum fluctuations.   The values of the counterterms are
independent of the external potential $V({\bf r})$.  Thus all ultraviolet 
divergences are removed by the same renormalizations that are 
required for the homogeneous Bose gas.

An alternative way to deal with ultraviolet divergences
is to replace the interatomic potential 
$(g + \delta g) \delta^3({\bf r})/2$ by a pseudopotential 
$g \delta^3({\bf r}) (\partial /\partial r) r/2$
\cite{Huang-Yang}.  Ultraviolet divergences can be avoided
by evaluating the partial derivative in the pseudopotential
at the appropriate stage of the calculation.  We find it simpler to 
introduce an ultraviolet cutoff and use the renormalization machinery 
of quantum field theory to remove the dependence on the cutoff. 

Most previous work on the density profile has been carried out within
the mean-field approximation.  In this approximation, 
quantum fluctuations are neglected.
The counterterms $\delta \mu$, $\delta g$, and $\delta \rho$,
which cancel ultraviolet divergences associated with 
those quantum fluctuations, can all be set to zero.
The field $\psi$ satisfies the time-dependent Gross-Pitaevskii equation,
which is the classical field equation for the action (\ref{action}):
\begin{equation}
i \hbar  {\partial \ \over \partial t} \psi \;+\;
\left(  {\hbar^2 \over 2 m} \mbox{\boldmath $\nabla$}^2 
	+ \mu - V({\bf r}) \right) \psi
	\;-\; {g \over 2} \left(\psi^\dagger \psi \right) \psi  
\;=\; 0 \,.
\label{GP}
\end{equation}
The ground state $| \Omega_\mu \rangle$
corresponds to a time-independent solution 
$\phi_0({\bf r})$ that can be chosen to be real-valued.
The mean field therefore satisfies
\begin{equation}
\left(  {\hbar^2 \over 2 m} \mbox{\boldmath $\nabla$}^2 
	+ \mu - V({\bf r}) \right) \phi_0({\bf r})
	\;-\; {g \over 2} \phi_0^3({\bf r})  
\;=\; 0 \,.
\label{GP-0}
\end{equation}
The number density (\ref{rho-QFT}) reduces to 
\begin{equation}
\rho({\bf r})	\;=\;  \phi_0^2({\bf r}) \,.
\label{rho-0}
\end{equation}
Thus the density profile in the mean-field 
approximation satisfies the Gross-Pitaevskii equation (\ref{GP-rho}).
The condensate in the mean-field 
approximation is
\begin{equation}
\langle \psi({\bf r}) \rangle \;=\; \phi_0({\bf r}) \,,
\label{vev-mfa}
\end{equation}
and it therefore satisfies  $\langle \psi \rangle = \sqrt{\rho}$.

The density profile is modified by quantum corrections.  
The corrections can be obtained by expanding the quantum field
around the mean field $\phi_0({\bf r})$, which satisfies (\ref{GP-0}):  
\begin{equation}
\psi({\bf r},t) 
\;=\; \phi_0({\bf r}) \;+\; \tilde{\psi}({\bf r},t) \,.
\label{psi-fluc}
\end{equation}
The expression (\ref{rho-QFT}) for the number density becomes
\begin{equation}
\rho({\bf r}) \;=\; \phi_0^2({\bf r})
\;+\; 2 \phi_0({\bf r}) {\rm Re} \langle \tilde{\psi}({\bf r}) \rangle
\;+\;\langle \tilde{\psi}^\dagger \tilde{\psi}({\bf r}) \rangle  
\;-\; \delta \rho \,.
\label{rho-fluc}
\end{equation}
The condensate differs from $\sqrt{\rho}$ because of the effects 
of quantum fluctuations:
\begin{equation}
\langle \psi({\bf r}) \rangle \;=\; \phi_0({\bf r})
\;+\;  \langle \tilde{\psi}({\bf r}) \rangle \,.
\label{vev-fluc}
\end{equation}

Having formulated our problem in terms of quantum field theory,
there are quantum fluctuations on all length scales ranging
from $L$, the length scale associated with variations in
$\rho({\bf r})$,
down to the inverse of the ultraviolet cutoff $\Lambda_{\rm UV}$.
The quantum fluctuations with length scales of order $L$
depend in detail on the shape of the potential $V({\bf r})$.  
For quantum fluctuations with length scales much smaller than $L$, 
the effects of variations in $V({\bf r})$ 
are negligible.  The short-distance quantum fluctuations 
therefore behave locally like those of a homogeneous Bose gas
with chemical potential $\mu - V({\bf r})$.
We will show that these fluctuations give the dominant corrections 
to the density profile. 

The one-loop quantum corrections to the number density can be obtained
by keeping the terms in the action that are quadratic 
in the fluctuating fields $\tilde{\psi}({\bf r},t)$.
If these fields are expanded in terms of normal modes,
the corrections (\ref{rho-fluc}) to $\rho({\bf r})$ can be expressed as 
a sum over the normal modes.  The contribution of an individual normal mode
to the number density scales like $1/L^3$.  This is 
negligible compared to the 
contribution from the mean field, 
which scales like $N/L^3$.  A significant contribution 
can only arise from summing over a large number of normal modes.
Normal modes with very short wavelengths approach a continuum 
and can be labeled by the wavevector ${\bf k}$. The contribution
to the density from such modes scales like $\int d^3k$.
The integral is ultraviolet divergent.  The 
ultraviolet divergence is proportional to $\Lambda_{\rm UV}^3$ and
is cancelled by the counterterm $\delta \rho$ in (\ref{rho-fluc}). 
After renormalization, modes with $k$ comparable to $\Lambda_{\rm UV}$
do not contribute to $\rho({\bf r})$.
Since the density of modes grows rapidly with $k$,
the dominant quantum corrections to $\rho({\bf r})$ come from the largest
values of $k$ whose effects are not removed by renormalization. 
To understand the scale of $k$ that dominates,
it is useful to recall some simple facts about the homogeneous Bose gas.

The properties of a homogeneous Bose gas with positive scattering length
$a$ and low number density $\rho$ are well understood.
The dimensionless quantity $\sqrt{\rho a^3}$ serves as an expansion 
parameter for the low-density expansion.  For example, 
the ground-state energy density, including the first quantum correction, 
is 
\begin{equation}
{\cal E} \;=\;  {2 \pi \hbar^2 a \rho^2 \over m}
\left( 1 \;+\; {128 \over 15 \sqrt{\pi}} \sqrt{\rho a^3} \right)  \,.
\label{E-Yang}
\end{equation}
The coefficient of $\sqrt{\rho a^3}$ in the quantum correction term
was first obtained by Lee and Yang \cite{Lee-Yang}.
The quasiparticle excitations of the system are
Bogoliubov modes, which are
plane waves with the dispersion relation
\begin{equation}
\epsilon(k) \;=\; {\hbar^2 k \sqrt{k^2 + \Lambda^2} \over 2 m} \,.
\label{epsilon}
\end{equation}
This dispersion relation
changes from linear in $k$ to quadratic at a  scale $\Lambda$ given by
\begin{equation}
\Lambda  \;=\; \sqrt{16 \pi a \rho} \,.
\label{Lambda}
\end{equation}
This is the scale of the wavenumber $k$ that dominates the quantum 
corrections to the energy density.  The one-loop quantum correction 
is the sum over normal modes of the zero-point energies
$\hbar \omega/2$, where $\omega$ 
is the angular frequency of the normal mode.
The contribution from large $k$ behaves like $\int d^3k \, \epsilon(k)/2$,
where $\epsilon(k)$ is the Bogoliubov dispersion relation given in 
(\ref{epsilon}).  This integral is ultraviolet divergent,
with the leading divergence proportional to 
$\hbar^2 \Lambda_{\rm UV}^5/m$.
This leading divergence and the subleading divergences can all
be removed by renormalization.  After renormalization, the integral is 
dominated by  the scale $\Lambda$ given in (\ref{Lambda}),
and it therefore scales like 
$\hbar^2 \Lambda^5/m  \sim  \hbar^2 \rho^{5/2} a^{5/2}/m$.
This estimate agrees with the explicit result given in (\ref{E-Yang}).
 
Generalizing to the case of a nonhomogeneous Bose gas, we can anticipate 
that the quantum corrections to the density profile $\rho({\bf r})$
will be dominated locally by modes with wavenumber $k$ on the order of 
$\sqrt{16 \pi a \rho({\bf r})}$.  
The contributions
from much shorter wavelengths are removed by renormalization.  
The contributions from much longer wavelengths are suppressed 
by phase space.
These modes can be approximated by a continuum as long as the corresponding 
wavelengths are much shorter than the scale $L$  
for significant variations in $\rho({\bf r})$:
\begin{equation}
\sqrt{16 \pi a \rho({\bf r})} \;\gg\; {2 \pi \over L} \,.
\label{Lambda-min}
\end{equation}
If this lower bound on the density is satisfied, then the 
methods of continuum quantum field theory can be used to calculate the
dominant quantum corrections.
The condition (\ref{Lambda-min}) is also necessary in order to calculate 
quantum corrections using a gradient expansion, 
which is an expansion in $1/(\Lambda L)$.
There is an upper bound on $\rho({\bf r})$ that must be satisfied
in order to allow perturbative calculations
in the  quantum field theory with the pointlike 
interaction in (\ref{action}).  This condition is that the 
scale $\Lambda$ must be much less than the maximum ultraviolet cutoff 
given in (\ref{LambdaUV-max}):
\begin{equation}
\sqrt{16 \pi a \rho({\bf r})} \;\ll\; {\pi \over 2 a} \,.
\label{Lambda-max}
\end{equation}
If this upper bound is not satisfied, then nonperturbative methods 
must be used to calculate the effects of quantum corrections.
The condition (\ref{Lambda-max}) is also necessary in order for
the effects of quantum fluctuations to be small enough 
to be treated as perturbative 
corrections to the mean-field approximation.
For example, in the case of a homogeneous Bose gas, the condition that
the quantum correction to the energy density, which is
given by the second term in (\ref{E-Yang}), is
small compared to the mean-field contribution is essentially 
identical to (\ref{Lambda-max}).
In our analysis of quantum corrections to the density profile,
we will assume that the number density is in the range specified 
by the inequalities (\ref{Lambda-min}) and (\ref{Lambda-max}).

\section{Perturbative Framework}
\label{sec:Cartesian}

In this Section, we present a general framework for carrying out perturbative 
calculations of the effects of quantum fluctuations around an arbitrary 
time-independent background $v({\bf r})$.  In Section \ref{sec:One-loop},
we will set $v$ equal to the mean field $\phi_0$ and use this framework to 
calculate one-loop corrections to the condensate and density profiles.
In Section \ref{sec:Self-consistent}, we set $v$ equal to the condensate
$\langle \psi \rangle$ and determine the self-consistent one-loop 
corrections to the equation for the density profile.

It is convenient to parameterize the quantum field
$\psi({\bf r},t)$ in terms of two real-valued quantum fields 
$\xi$ and $\eta$ that describe 
quantum fluctuations around an arbitrary 
time-independent background $v({\bf r})$:
\begin{equation}
\psi({\bf r},t) \;=\;  v({\bf r})
\;+\; { \xi({\bf r},t) + i \eta({\bf r},t) \over \sqrt{2} } \,.
\label{psi-cart}
\end{equation}
We will refer to this as the {\it Cartesian parameterization} 
of the quantum field.  
An alternative field parameterization is considered 
in Section \ref{sec:Polar}.
If the phase of $\psi$ is chosen so that 
$v$ is real-valued, the condensate profile is
\begin{equation}
\langle \psi({\bf r}) \rangle \;=\;  v({\bf r})
\;+\; {1 \over \sqrt{2} } \langle  \xi({\bf r}) \rangle \,.
\label{vev-cart}
\end{equation}
The number density is 
\begin{equation}
\rho({\bf r}) \;=\;  v^2({\bf r})
\;+\; \sqrt{2}  v({\bf r}) \big\langle  \xi({\bf r}) \big\rangle 
\;+\; {1 \over 2} \left\langle \xi^2({\bf r}) \right\rangle
\;+\; {1 \over 2} \left\langle \eta^2({\bf r}) \right\rangle
\;-\; \delta \rho \,.
\label{rho-cart}
\end{equation}

Inserting the field parameterization (\ref{psi-cart})
into the action (\ref{action}) and expanding in powers of the quantum fields
$\xi$ and $\eta$, it becomes
\begin{eqnarray}
S[\psi] \;=\;   S[v] &+& \int dt \int d^3r \Bigg\{ 
\sqrt{2} T \xi
\;+\; {1 \over 2} \left( \eta \dot{\xi} - \xi \dot{\eta} \right)
\;+\; {1 \over 4m} \xi \left( \mbox{\boldmath $\nabla$}^2 
				- \Lambda^2 + X \right) \xi
\nonumber \\
&& \;+\; {1 \over 4m} \eta \left( \mbox{\boldmath $\nabla$}^2 + Y \right) \eta
\;+\;  {1 \over \sqrt{2}} Z \xi \left( \xi^2 + \eta^2 \right) 
\;-\;  {g + \delta g \over 16} \left( \xi^2 + \eta^2 \right)^2 
\Bigg\}\,,
\label{S-cart}
\end{eqnarray}
where $\dot{f} \equiv {\partial \ \over \partial t} f$ and 
$T$, $X$, $Y$, and $Z$ are external sources that depend on $v$:
\begin{eqnarray}
T({\bf r}) &=& 
\left[ (\mu + \delta \mu) - V({\bf r}) 
 	- {g + \delta g \over 2} v^2({\bf r}) \right] v({\bf r})
 \;+\; {1 \over 2 m} \mbox{\boldmath $\nabla$}^2 v({\bf r}) \,,
\label{T-def}
\\
X({\bf r}) &=& \Lambda^2
	\;+\; 2 m \left[ (\mu + \delta \mu) - V({\bf r}) 
	- {3 (g + \delta g) \over 2} v^2({\bf r}) \right] \,,
\label{X-def}
\\
Y({\bf r}) &=& 2m \left[ (\mu + \delta \mu)  - V({\bf r}) 
	- {g + \delta g \over 2} v^2({\bf r}) \right] \,,
\label{Y-def}
\\
Z({\bf r}) &=& - {g + \delta g \over 2} v({\bf r}) \,.
\label{Z-def}
\end{eqnarray}
We have set $\hbar = 1$ in the action.  
Dimensional analysis can be used to reinsert the factors of
$\hbar$ at the end of the calculation. 
The parameter $\Lambda$ appears both in the source $X$ and explicitly in the 
$\xi^2$ term in  the action,
and cancels between them.  The arbitrariness of this
parameter can be exploited to simplify calculations.

To organize the quantum corrections into a loop expansion,
we separate the terms in the action that depend on $\xi$ and $\eta$
into a free part and an interaction part:
\begin{equation}
S[\psi] \;=\;  S[v] 
\;+\; S_{\rm free}[\xi,\eta] \;+\; S_{\rm int}[v,\xi,\eta]  \,.
\label{S-decomp}
\end{equation}
The free part of the action is 
\begin{equation}
S_{\rm free}[\xi,\eta] \;=\;  \int dt \int d^3r \Bigg\{ 
{1 \over 2} \left( \eta \dot{\xi} - \xi \dot{\eta} \right)
\;+\; {1 \over 4 m} \xi ( \mbox{\boldmath $\nabla$}^2 - \Lambda^2 ) \xi
\;+\; {1 \over 4 m} \eta \mbox{\boldmath $\nabla$}^2 \eta \Bigg\} \,.
\label{S-free}
\end{equation}
This action describes Bogoliubov modes with the dispersion relation
(\ref{epsilon}), where $\Lambda$ is now an adjustable parameter.  
The Fourier transform of the propagator for the fields 
$\xi$ and $\eta$ is a $2 \times 2$ matrix:
\begin{equation}
\left( 
\begin{array}{cc}
	D^{\xi \xi}({\bf k},\omega)  & D^{\xi \eta}({\bf k},\omega) 	\\
	D^{\eta \xi}({\bf k},\omega) & D^{\eta \eta}({\bf k},\omega) 
	\end{array}
\right)
\;=\;
{1 \over  \omega^2 - \epsilon^2(k) + i 0^+}
\left( 
\begin{array}{cc}
	k^2/(2 m) & - i \omega	\\
	i \omega    & 2 m \epsilon^2(k)/k^2
	\end{array}
\right) \,,
\label{propagator}
\end{equation}
where ${\bf k}$ is the wavevector and $\omega$ is the frequency.
The diagonal elements of the propagator matrix (\ref{propagator})
are represented by solid lines for $\xi$ and dashed lines for $\eta$,
as illustrated in Figures 1(a) and 1(b).  
The off-diagonal elements are represented 
by a line that is half solid and half dotted, as in Figure 1(c).
All the remaining terms in the action (\ref{S-cart})
are treated as interactions:
\begin{eqnarray}
S_{\rm int}[v,\xi,\eta] &=& \int dt \int d^3r \Bigg\{ 
\sqrt{2} T \xi
\;+\; {1 \over 4m} X \xi^2
\;+\; {1 \over 4m} Y \eta^2
\nonumber \\
&&  \qquad \qquad
\;+\;  {1 \over \sqrt{2}} Z \xi \left( \xi^2 + \eta^2 \right) 
\;-\;  {g + \delta g \over 16} \left( \xi^2 + \eta^2 \right)^2 
\Bigg\} \,.
\label{Sint-cart}
\end{eqnarray}
They include interactions with the external sources $T$, $X$, $Y$, 
and $Z$ as well as the 4-point coupling $g + \delta g$.
The sources are represented by dots labeled by the appropriate letter,
as illustrated in Figure 2. The 4-point couplings are represented by 
points that connect four lines. 

It is possible to diagonalize the propagator matrix (\ref{propagator})
by applying a Bogoliubov transformation to the fields $\xi$ and $\eta$.
However, such a transformation makes the interaction terms 
in the action significantly more complicated and increases
the number of diagrams that contribute to most quantities.
For explicit calculations, it is more economical to 
minimize the number of diagrams.  We therefore prefer to use a 
propagator matrix with off-diagonal elements.

In the case of a homogeneous Bose gas, the trapping potential
$V$ is zero and we can choose the background field $v$ to be a constant,
independent of ${\bf r}$.  If we choose 
$v^2 = 2 (\mu + \delta \mu)/(g + \delta g)$ and 
$\Lambda^2 = 2 m (g + \delta g) v^2$, then the sources $T$, $X$, and $Y$
in (\ref{Sint-cart}) vanish and the interactions reduce to
three-point couplings and four-point couplings.
Such a perturbative framework has
recently been used to reproduce the classic
one-loop corrections to the thermodynamic properties of a 
homogeneous Bose gas \cite{Ravndal}.

The leading quantum corrections to the ground-state expectation values
in (\ref{vev-cart}) and (\ref{rho-cart})
are given by one-loop Feynman diagrams.   Examples of one-loop diagrams that 
contribute to $\langle \xi^2 \rangle$
and $\langle \eta^2 \rangle$ are shown in Figures 3 and 4, respectively. 
The dot on the left side of each diagram represents the operator $\xi^2$,
which creates two solid lines, or 
the operator $\eta^2$, which creates two dashed lines.
The lines form a loop that can include any number of insertions 
of the sources $X$ and $Y$.  It is convenient to introduce the notation
$\langle \ldots \rangle_{XY}$ for the expectation value of an 
operator in the presence of the sources $X$ and $Y$,
but with no other self-interactions for the quantum fields.
The sum of all one-loop diagrams for $\langle \xi^2 \rangle$
and $\langle \eta^2 \rangle$ can then be represented as
\begin{eqnarray}
\left\langle \xi^2({\bf r}) \right\rangle \Bigg|_{\rm 1-loop} &=& 
\left\langle \xi^2({\bf r}) \right\rangle_{X,Y} \,,
\label{xi-ms}
\\
\left\langle \eta^2({\bf r}) \right\rangle \Bigg|_{\rm 1-loop}&=& 
\left\langle \eta^2({\bf r}) \right\rangle_{X,Y} \,.
\label{eta-ms}
\end{eqnarray}
The advantage of this notation is that
the expectation value $\langle \xi \rangle $ at one-loop order can also be 
expressed succinctly in terms of $\langle \xi^2 \rangle_{X,Y}$
and $\langle \eta^2 \rangle_{X,Y}$.
Examples of diagrams that contribute to $\langle \xi \rangle$
are shown in Figures 5, 6, and 7.  The operator $\xi$ creates a single 
solid line.  In the diagrams of Figure 5, the $\xi$ field propagates to a 
source $T$, where it is annihilated.  In the diagrams of Figures 6 and 7,
it propagates to a source $Z$, which creates 
a pair of solid lines or dashed lines that form a loop.  
In all of these diagrams,
the $\xi$ propagator and the propagators inside the loop can
include any number of insertions of the sources $X$ and $Y$.
The sum of all such diagrams can be expressed as
\begin{eqnarray}
\big\langle \xi({\bf r}) \big\rangle  \Bigg|_{\rm 1-loop} 
&=& - \sqrt{2} \int d^3r' 
\left( \int dt' \, D^{\xi \xi}_{X,Y}({\bf r},{\bf r}',t') \right)
\nonumber \\ 
&& \qquad
\times \left( T({\bf r}') 
\;+\; {3 \over 2}  Z({\bf r}') 
	\left\langle \xi^2({\bf r}') \right\rangle_{X,Y}
\;+\; {1 \over 2}  Z({\bf r}') 
	\left\langle \eta^2({\bf r}') \right\rangle_{X,Y}
\right) \,,
\label{xi-m}
\end{eqnarray}
where $D^{\xi \xi}_{X,Y}$ is the diagonal component of the propagator
for $\xi$ in the presence of the sources $X$ and $Y$.

The quantities $\langle \xi^2 \rangle_{X,Y}$ and 
$\langle \eta^2 \rangle_{X,Y}$ in (\ref{xi-ms}), (\ref{eta-ms}),
and (\ref{xi-m}) are functionals of the sources $X$ and $Y$.
These functionals include terms of arbitrarily high orders in 
$X$ and $Y$. 
They are nonlocal because the loop diagrams involve an
integral over the positions  of the sources $X$ and $Y$.
After renormalization,
these integrals are dominated by wavelengths of order $2 \pi/\Lambda$,
while the sources vary significantly only over much larger 
distances of order $L$.  It is therefore reasonable to expand the sources
$X({\bf r}')$ and $Y({\bf r}')$ as Taylor series around the point ${\bf r}$.  
This reduces the expressions for  $\langle \xi^2 \rangle_{X,Y}$ and 
$\langle \eta^2 \rangle_{X,Y}$ to an infinite sum of local 
quantities involving $X$, $Y$, and their derivatives at the point ${\bf r}$:
\begin{eqnarray}
\left\langle \xi^2({\bf r}) \right\rangle_{X,Y}  
&=& a_0  \;+\; a_1 \, X({\bf r}) \;+\; a_2 \, Y({\bf r})
\;+\; a_3 \mbox{\boldmath $\nabla$}^2 X({\bf r}) 
\nonumber \\
&&
\;+\; a_4 \, X^2({\bf r})
\;+\; a_5 ( \mbox{\boldmath $\nabla$} X)^2({\bf r})
\;+\; \ldots  \,,
\label{xi-ms:DE}
\\
\left\langle \eta^2({\bf r}) \right\rangle_{X,Y}  
&=& b_0  \;+\; b_1 \, X({\bf r}) \;+\; b_2 \, Y({\bf r})
\;+\; b_3 \mbox{\boldmath $\nabla$}^2 X({\bf r}) 
\nonumber \\
&&
\;+\; b_4 \, X^2({\bf r})
\;+\; b_5 ( \mbox{\boldmath $\nabla$} X )^2({\bf r}) 
\;+\; \ldots  \,.
\label{eta-ms:DE}
\end{eqnarray}
The terms on the right sides of (\ref{xi-ms:DE}) and (\ref{eta-ms:DE})
include all possible rotationally invariant combinations of 
$X$ and $Y$ and their derivatives.
The constraint of rotational
invariance arises from the
rotational invariance of the free action (\ref{S-free}).
In (\ref{xi-ms:DE}) and (\ref{eta-ms:DE}), we have shown explicitly 
only those terms that will ultimately be needed to calculate
the quantum corrections to the density profile.

The coefficients $a_i$ and $b_i$ in (\ref{xi-ms:DE}) 
and (\ref{eta-ms:DE}) can be reduced to integrals over a 
wave-vector ${\bf k}$, as illustrated by the explicit calculation
of a diagram presented in Appendix \ref{app:Diagram}.
Having expanded the sources as Taylor series around ${\bf r}$,
the only scale in the integrand is $\Lambda$.
By dimensional analysis, a convergent integral must have the form of the 
appropriate power of $\Lambda$ multiplied by a numerical coefficient.
However some of the integrals have infrared or ultraviolet divergences,
and thus require infrared or ultraviolet cutoffs. 
The ultraviolet divergences either cancel in  quantities 
such as $\rho({\bf r})$ and $\langle \psi({\bf r}) \rangle$
or they are removed by renormalization.
Infrared divergences reflect a failure of the assumption that
the sources can be expanded in a Taylor series inside the loop integral.  
If these divergences do not cancel, it simply 
indicates a breakdown of the gradient expansion due to the sensitivity 
of the quantum corrections to nonlocal effects involving the length scale $L$.
 
The propagator factor $\int dt' D^{\xi \xi}_{X,Y}({\bf r},{\bf r}',t')$ 
in (\ref{xi-m}) can be expanded
in powers of $X$ and its derivatives at the point ${\bf r}$.
The dependence on the source $Y$ is removed by 
the integration over $t'$, which  
corresponds 
to evaluating the Fourier-transformed propagator at $\omega = 0$.
Since the off-diagonal components of the propagator (\ref{propagator})
vanish at zero frequency, the source $Y$ does not contribute.
Examples of diagrams that contribute to $\int dt' \, D^{\xi \xi}_{X,Y}$
are shown in Figure 8.  The contribution from the first diagram
is given by the upper-left component of the propagator matrix in 
(\ref{propagator}):
\begin{equation}
\int dt' \, D^{\xi \xi}({\bf r},{\bf r}',t') 
\;=\;  - 2m \int {d^3k \over (2 \pi)^3} 
	e^{-i {\bf k} \cdot ({\bf r} - {\bf r}')}
{1 \over k^2 + \Lambda^2} \,.
\label{Dxixi}
\end{equation}
The other diagrams in Figure 8 involve integrals over the positions
${\bf r}''$ of sources $X({\bf r}'')$.  In coordinate space, 
the propagator factor
(\ref{Dxixi}) falls exponentially when $|{\bf r} - {\bf r}'|$ exceeds 
$1/\Lambda$.  If we assume that the source $X$ varies significantly 
only over a much greater length scale $L$, then we can expand 
$X({\bf r}'')$ as a Taylor series around the point ${\bf r}'' = {\bf r}$.
The function $\int dt' \, D^{\xi \xi}_{X,Y}({\bf r},{\bf r}',t')$
can then be expressed in terms of $X({\bf r})$ and its derivatives 
at the point ${\bf r}$.  The terms coming from the diagrams in 
Figure 8 include
\begin{eqnarray}
\int dt' \, D^{\xi \xi}_{X,Y}({\bf r},{\bf r}',t') 
&=&  - 2m \int {d^3k \over (2 \pi)^3} 
	e^{-i {\bf k} \cdot ({\bf r} - {\bf r}')}
\Bigg\{ {1 \over k^2 + \Lambda^2} 
\;+\; X({\bf r}) {1 \over (k^2 + \Lambda^2)^2}
\nonumber \\
&&
\;-\; 2 i \nabla_i X({\bf r}) {k^i \over (k^2 + \Lambda^2)^3}
\;+\;  \nabla_i \nabla_j X({\bf r}) 
	\left[ {\delta^{ij} \over (k^2 + \Lambda^2)^3}
	- 4 {k^i k^j \over (k^2 + \Lambda^2)^4} \right]
\nonumber \\
&&
\;+\; 2 \nabla_i X \nabla_j X({\bf r}) 
	\left[ {\delta^{ij} \over (k^2 + \Lambda^2)^4}
	- 6 {k^i k^j \over (k^2 + \Lambda^2)^5} \right]
\;+\; \ldots \Bigg\} \,.
\label{Dxixi-XY}
\end{eqnarray}
The complete expression involves all possible powers of $X$ and 
gradients of $X$. 

The result (\ref{Dxixi-XY}) can be used to express
$\langle \psi({\bf r}) \rangle$ and $\rho({\bf r})$ as an expansion in 
powers of $X$ and $Y$ and their derivatives at the point $({\bf r})$.
In the expression for $\langle \xi({\bf r}) \rangle$ in (\ref{xi-m}), 
the propagator factor is integrated against a function 
$f({\bf r}')$ that depends on the sources $T$, $Z$, $X$, and $Y$.
The integral can be evaluated by expanding $f({\bf r}')$ as a Taylor series
around the point ${\bf r}' = {\bf r}$.  
Using the expression (\ref{Dxixi-XY}) for the propagator factor,
we can evaluate the integral over ${\bf r}'$.
The resulting expression for the integral includes the terms
\begin{eqnarray}
&&\int d^3 r' 
\left( \int dt' \, D^{\xi \xi}_{X,Y}({\bf r},{\bf r}',t') \right)
	f({\bf r}')
\nonumber \\ 
&& \qquad
\;=\; - 2m \Bigg\{ \left[ {1 \over \Lambda^2} 
+ {1 \over \Lambda^4} X({\bf r})
+ {1 \over \Lambda^6} \mbox{\boldmath $\nabla$}^2 X({\bf r})
+ {2 \over \Lambda^8} (\mbox{\boldmath $\nabla$}X)^2({\bf r})
	\right] f({\bf r})
\nonumber \\ 
&&  \qquad \qquad \qquad \qquad
\;+\; {2 \over \Lambda^6} \mbox{\boldmath $\nabla$} X({\bf r})
	\cdot \mbox{\boldmath $\nabla$} f({\bf r})
\;+\;  {1 \over \Lambda^4} \mbox{\boldmath $\nabla$}^2 f({\bf r})
\;+\; \ldots \Bigg\}\,.
\label{int-Df}
\end{eqnarray}
Applying this formula to the integral in (\ref{xi-m}) and using the
expansions (\ref{xi-ms:DE}) and (\ref{eta-ms:DE})
for $\langle \xi^2 \rangle$ and $\langle \eta^2 \rangle$, we obtain
an expansion for $\langle \xi \rangle$ in powers of $X$, $Y$, 
and their derivatives. 
Inserting the expansions for $\langle \xi \rangle$,
$\langle \xi^2 \rangle$, and $\langle \eta^2 \rangle$
into (\ref{vev-cart}) and (\ref{rho-cart}), we obtain expansions 
for the condensate and the density in powers of $X$, $Y$, 
and their derivatives.

\section{One-loop calculation}
\label{sec:One-loop}

In this section, we calculate the one-loop quantum corrections 
to the density profile $\rho({\bf r})$ and  
to the condensate profile $\langle \psi({\bf r}) \rangle$
to second order in the gradient expansion.
The appropriate choice for the background field $v$ is the mean field
$\phi_0$,
which satisfies (\ref{GP-0}):
\begin{equation}
v({\bf r}) \;=\;  \phi_0({\bf r}) \,.
\label{psi-1loop}
\end{equation}
The quantum fields $\xi$ and $\eta$ in (\ref{psi-cart}) describe 
quantum fluctuations around the mean field.
The condensate profile (\ref{vev-cart}) reduces to 
\begin{equation}
\langle \psi({\bf r}) \rangle \;=\; \phi_0({\bf r})
\;+\; {1 \over \sqrt{2}} \langle \xi({\bf r}) \rangle \,,
\label{vev-1loop}
\end{equation}
while the number density (\ref{rho-cart}) reduces to 
\begin{equation}
\rho({\bf r}) \;=\; \phi_0^2({\bf r})
\;+\; \sqrt{2} \phi_0({\bf r}) \langle \xi({\bf r}) \rangle 
\;+\; {1 \over 2} \left\langle \xi^2({\bf r}) \right\rangle 
\;+\; {1 \over 2} \left\langle \eta^2({\bf r}) \right\rangle 
\;-\; \delta \rho \,.
\label{rho-1loop}
\end{equation}
The fact that the mean field $\phi_0$ satisfies the 
classical equation (\ref{GP-0}) can be used to simplify the expressions 
(\ref{T-def})--(\ref{Z-def}) for the sources.  We can also drop the
counterterms $\delta \mu$ and $\delta g$ in the sources
$X$, $Y$, and $Z$.  These sources appear only in diagrams that are 
at least first order in the loop expansion.  The counterterms appearing 
in these sources are therefore needed only to cancel ultraviolet 
divergences that arise at second order or higher in the loop expansion.
Thus the sources can be simplified to 
\begin{eqnarray}
T({\bf r}) &=&  
\delta \mu \,  \phi_0({\bf r})
\;-\; {\delta g \over 2}  \phi_0^3({\bf r}) \,,
\label{T-1loop}
\\
X({\bf r}) &=& 
\Lambda^2 \;-\; 2 m g \phi_0^2({\bf r})
\;-\; {\mbox{\boldmath $\nabla$}^2 \phi_0 \over \phi_0}({\bf r}) \,,
\label{X-1loop}
\\
Y({\bf r}) &=&
- {\mbox{\boldmath $\nabla$}^2 \phi_0 \over \phi_0}({\bf r}) \,,
\label{Y-1loop}
\\
Z({\bf r}) &=& - {g \over 2} \phi_0({\bf r}) \,.
\label{Z-1loop}
\end{eqnarray}

The expressions (\ref{vev-1loop}) and (\ref{rho-1loop}) for the condensate
and the density are nonlocal functionals of the mean field $\phi_0$. 
If $\langle \psi({\bf r}) \rangle$ and $\rho({\bf r})$ are expanded
in powers of the sources $X$ and $Y$ and their derivatives 
at the point ${\bf r}$, the expansions include
infinitely many terms.  They can be reduced to local 
functionals of $\phi_0$ by consistently truncating the expansions.
We will reduce (\ref{vev-1loop}) and (\ref{rho-1loop})
to local equations at a specific point ${\bf r}_0$ by
(a) choosing a specific value for the arbitrary parameter $\Lambda$
and (b) truncating the equations
at second order in the gradient expansion.  
Note that the source $Y$ in (\ref{Y-1loop}) is already second order in the 
gradient expansion.  Thus if we truncate
the equations at second order  in the gradient expansion,
we need only include terms up to first order in $Y$
and we can omit all derivatives of $Y$.
We also need only include terms up to first order in 
$\mbox{\boldmath $\nabla$}^2 X$ and up to second order in
$\mbox{\boldmath $\nabla$} X$.  However, we still must include 
all possible powers of $X$.

In order to reduce the expansions for $\langle \psi \rangle$ and $\rho$
to a finite number of terms, we will choose $\Lambda$ so that 
$X({\bf r}_0)$ is second order in the gradient expansion at a specific 
point ${\bf r}_0$.  If we evaluate $\langle \psi \rangle$ and $\rho$
at the point ${\bf r}_0$
and then truncate them at second order in gradients of $\phi_0$,
the resulting expressions for $\langle \psi({\bf r}_0) \rangle$
and $\rho({\bf r}_0)$ are algebraic functions of
$\phi_0$, $\mbox{\boldmath $\nabla$} \phi_0$,
and $\mbox{\boldmath $\nabla$}^2 \phi_0$ evaluated at the point ${\bf r}_0$.
Since we could have chosen any particular point for ${\bf r}_0$,
these algebraic relations must hold at any point ${\bf r}$.
The most convenient choice for $\Lambda$ 
is the wavenumber that appears in the Bogoliubov 
dispersion relation (\ref{epsilon}) for a homogeneous gas with
number density $\phi_0^2({\bf r}_0)$:
\begin{equation}
\Lambda^2 \;=\; 2 m g \phi_0^2({\bf r}_0) \,.
\label{Lambda-1loop}
\end{equation}
The source $X$ and its derivatives at the point ${\bf r}_0$
then reduce to 
\begin{eqnarray}
X({\bf r}_0) &=& - 
{\mbox{\boldmath $\nabla$}^2 \phi_0 \over \phi_0}({\bf r_0}) \,,
\label{X-1loop:r0}
\\
\mbox{\boldmath $\nabla$} X({\bf r}_0) &=& 	
- 4 m g \, \phi_0 \mbox{\boldmath $\nabla$} \phi_0({\bf r}_0) \,,
\label{dX-1loop:r0}
\\
\mbox{\boldmath $\nabla$}^2 X({\bf r}_0) &=& 
- 4 m g \left[ \phi_0 \, \mbox{\boldmath $\nabla$}^2 \phi_0
	+ (\mbox{\boldmath $\nabla$} \phi_0)^2 \right]({\bf r}_0) \,.
\label{ddX-1loop:r0}
\end{eqnarray}

We proceed to calculate the one-loop correction to the condensate
$\langle \psi({\bf r}) \rangle$, which is given by (\ref{xi-m}).
Inserting (\ref{xi-ms:DE}) and (\ref{eta-ms:DE}) into (\ref{xi-m})
and using (\ref{int-Df}) to evaluate the integral over ${\bf r}'$,
we obtain an expansion for $\langle \xi \rangle$ in powers of the sources
and their derivatives.
Inserting the expressions (\ref{T-1loop}),  (\ref{Y-1loop}),
(\ref{Z-1loop}), and (\ref{X-1loop:r0})-(\ref{ddX-1loop:r0}) for the
sources at the point ${\bf r}_0$, we obtain
\begin{eqnarray}
\langle \xi({\bf r}_0) \rangle &=&
- \sqrt{2} m g \phi_0 
\Bigg\{
\left[ {3a_0 + b_0 \over 2}  - 2 {\delta \mu \over g} 
	+ {\delta g \over g} \phi_0^2 \right] 
	{1 \over \Lambda^2}
\nonumber \\
&& 
\;-\; \left[ \left( 3 a_0 + b_0 - 4 {\delta \mu \over g} \right)
		{1 \over \Lambda^4}
	+ {9a_1 + 3 b_1 + 3a_2 + b_2 \over 2} {1 \over \Lambda^2}
	+ (3 a_3 + b_3) \right]
	{\mbox{\boldmath $\nabla$}^2 \phi_0 \over \phi_0}
\nonumber \\
&& 
\;+\; \Bigg[ \left( 3 a_0 + b_0 - 4 {\delta \mu \over g} \right)
		{1 \over \Lambda^4}
	+ (3 a_1 + b_1) {1 \over \Lambda^2}
\nonumber \\
&& \qquad \qquad
	- (3 a_3 + b_3 - 12 a_4 - 4 b_4) 
	+ 2 ( 3 a_5 + b_5 ) \Lambda^2 \Bigg]
	{(\mbox{\boldmath $\nabla$} \phi_0)^2 \over \phi_0^2} 
\Bigg\} \,.
\label{xi-1loop}
\end{eqnarray}
The coefficients $a_i$ and $b_i$ are given in Appendix \ref{app:Coefficients}. 
The coefficients $a_0$ and $b_0$ are cubically ultraviolet divergent, 
while $a_1$ and $b_1$ are linearly divergent. 
The divergences are cancelled by taking the counterterms 
$\delta \mu$ and $\delta g$ to have the values
\begin{eqnarray}
\delta \mu &=& 
{1 \over 12 \pi^2} g \Lambda_{\rm UV}^3  \,, 
\label{delmu}
\\
\delta g &=& 
{1 \over 4 \pi^2}  \left( m g \Lambda_{\rm UV} \right) g \,.
\label{delg}
\end{eqnarray}
The counterterm $\delta g$ in (\ref{delg}) agrees with that obtained by
expanding (\ref{g-bare}) to first order in $m g \Lambda_{\rm UV}$.
Using the results for $a_i$ and $b_i$ in Appendix \ref{app:Coefficients} 
and the value of $\Lambda$ given in (\ref{Lambda-1loop}), 
the condensate at the point ${\bf r}_0$ reduces to
\begin{eqnarray}
\langle \psi \rangle
\;=\; \phi_0
\Bigg\{ 1 
\;-\; {5 \over 48 \pi^2} (2 m g)^{3/2} \phi_0
\;-\; {1 \over 16 \pi^2} \sqrt{2 m g}
\left[
\left( {49 \over 18} - {5 \over 24} 
		\log {8 m g \phi_0^2  \over \lambda_{\rm IR}^2} \right)
	{\mbox{\boldmath $\nabla$}^2 \phi_0 \over \phi_0^2}
\right.
\nonumber \\
\left.
\;+\; \left( {29 \over 9} - {1 \over 16} 
		\log {8 m g \phi_0^2 \over \lambda_{\rm IR}^2} \right)
	{(\mbox{\boldmath $\nabla$} \phi_0)^2 \over \phi_0^3} 
\right] 
\Bigg\} \,,
\label{vev-1loop:1}
\end{eqnarray}
where $\lambda_{\rm IR}$ is an infrared cutoff.
The logarithmic infrared divergences arise from the coefficients 
$b_2$, $b_3$, and $b_5$.  The divergences indicate that the gradient 
expansion for  the condensate breaks down at second order.
Thus we can obtain a local expression for the condensate 
only to leading order in the gradient expansion. 
Keeping only the first correction term in (\ref{vev-1loop:1}), the result is 
\begin{equation}
\langle \psi({\bf r}) \rangle
\;=\; \phi_0({\bf r})
\left[ 1 \;-\; {5 \over 48 \pi^2} (2 m g)^{3/2} \phi_0({\bf r}) \right] \,.
\label{vev-1loop:2}
\end{equation}
We derived this equation at the point ${\bf r}_0$ defined by  
our choice (\ref{Lambda-1loop}) for the arbitrary parameter $\Lambda$.
However, our final result for $\langle \psi({\bf r}_0) \rangle$ 
is an algebraic expression in terms of $\phi_0({\bf r}_0)$.
Since we could have chosen any particular point for ${\bf r}_0$,
that algebraic expression must be valid at any point ${\bf r}$.

We next calculate the one-loop corrections to the density, 
which is given by (\ref{rho-1loop}).
The expression for $\langle \xi \rangle$ at the point ${\bf r}_0$
is given by (\ref{xi-1loop}).  The corresponding expressions for 
$\langle \xi^2 \rangle$  and $\langle \eta^2 \rangle$ are obtained 
by inserting the expressions (\ref{Y-1loop}) and 
(\ref{X-1loop:r0})-(\ref{ddX-1loop:r0}) for the
sources at the point ${\bf r}_0$ into (\ref{xi-ms:DE}) and 
(\ref{eta-ms:DE}). The resulting expression for the density at the 
point ${\bf r}_0$ is
\begin{eqnarray}
\rho({\bf r}_0) \;=\;
\phi_0^2 
&-& \left[ a_0 + \delta \rho  - 2 {\delta \mu \over g} 
	+ {\delta g \over g} \phi_0^2 \right] 
\nonumber \\
&+& \left[ \left( 3 a_0 + b_0 - 4 {\delta \mu \over g} \right)
		{1 \over \Lambda^2}
	+ ( 4 a_1 + b_1 + a_2 )
	+ 2 a_3 \Lambda^2 \right]
	{\mbox{\boldmath $\nabla$}^2 \phi_0 \over \phi_0}
\nonumber \\
&-& \Bigg[ \left( 3 a_0 + b_0 - 4 {\delta \mu \over g} \right)
		{1 \over \Lambda^2}
	+ ( 3 a_1 + b_1 )
\nonumber \\
&& \qquad
	- 2 (a_3 - 6 a_4 - 2 b_4) \Lambda^2
	+ 4 a_5 \Lambda^4 \Bigg]
	{(\mbox{\boldmath $\nabla$} \phi_0)^2 \over \phi_0^2} \,.
\label{rho-1loop:1}
\end{eqnarray}
After using the expressions (\ref{delmu}) and (\ref{delg}) 
for the counterterms $\delta \mu$ and $\delta g$, the only remaining
ultraviolet divergence is a cubic divergence that can be
cancelled by choosing the density counterterm to be
\begin{equation}
\delta \rho \;=\; 
{1 \over 12 \pi^2} \Lambda_{\rm UV}^3  \,.
\label{delrho}
\end{equation}
The infrared divergent coefficients $b_2$, $b_3$, and $b_5$ have 
cancelled in the expression (\ref{rho-1loop:1}) for the number density.
Thus the density has a well-defined gradient expansion through
second order, in contrast to the condensate. 
Our final expression for the number density, 
including one-loop quantum corrections, is
\begin{eqnarray}
\rho({\bf r}) \;=\; \phi_0^2({\bf r}) 
\Bigg\{ 1
&-& {1 \over 6 \pi^2} (2 m g)^{3/2} \phi_0({\bf r})
\nonumber \\
&-& {1 \over 16 \pi^2} \sqrt{2 m g}
\left[ {41 \over 9} 
	{\mbox{\boldmath $\nabla$}^2 \phi_0 \over \phi_0^2}({\bf r})
\;+\; {113 \over 18} 
	{(\mbox{\boldmath $\nabla$} \phi_0)^2 \over \phi_0^3}({\bf r}) 
\right] 
\Bigg\} \,.
\label{rho-1loop:2}
\end{eqnarray}
We derived this equation at the specific point ${\bf r}_0$.
However, our final expressions for $\rho({\bf r}_0)$ is an algebraic 
expression involving $\phi_0$, $\mbox{\boldmath $\nabla$} \phi_0$,
and $\mbox{\boldmath $\nabla$}^2 \phi_0$ evaluated at the point ${\bf r}_0$.
Since we could have chosen any particular point for ${\bf r}_0$,
these algebraic relations must hold at any point ${\bf r}$.

Combining (\ref{vev-1loop:2}) and (\ref{rho-1loop:2}), we obtain a local
expression for the condensate in terms of the density that is correct 
to leading order in the gradient expansion: 
\begin{equation}
\langle \psi({\bf r}) \rangle
\;=\; \sqrt{\rho}({\bf r})
\left[ 1 \;-\; {1 \over 48 \pi^2} (2 m g)^{3/2} \sqrt{\rho}({\bf r}) 
\right] \,.
\label{vev-rho}
\end{equation}
This agrees with a result obtained recently by Timmermans,
Tommasini, and Huang \cite{T-T-H}. 

The choice (\ref{Lambda-1loop}) for $\Lambda$ is not unique.
Any choice that makes $X({\bf r}_0)$ second order in the gradient expansion 
will be equally acceptable and must give the same final answer.
For example, we could have chosen
\begin{equation}
\Lambda^2 \;=\; 2 m g \phi_0^2({\bf r}_0) 
\;+\; {\mbox{\boldmath $\nabla$}^2 \phi_0 \over \phi_0}({\bf r}_0)\,.
\label{Lambda-alt}
\end{equation}
In that case, (\ref{X-1loop:r0}) would be replaced by $X({\bf r}_0) = 0$.
Following the effects of this change through the calculation,
we find that the coefficient $4 a_1$ of 
$\mbox{\boldmath $\nabla$}^2 \phi_0/\phi_0$ in (\ref{rho-1loop:1})
is replaced by $3 a_1$.  
However, the term $-a_0$ in (\ref{rho-1loop:1}) depends on $\Lambda$,
which is given in (\ref{Lambda-alt}). When this term is expanded
in powers of gradients of $\phi_0$, it generates additional terms 
proportional  to $\mbox{\boldmath $\nabla$}^2 \phi_0/\phi_0$
that precisely cancel the change in (\ref{rho-1loop:1}).
Thus we recover the same final result (\ref{rho-1loop:2}).

Note that the counterterms (\ref{delmu}), (\ref{delg}), and (\ref{delrho})
do not depend on the potential $V({\bf r})$.  
Thus the ultraviolet divergences in one-loop diagrams
are removed by the same
renormalizations that are required for a homogeneous Bose gas.

\section{Self-consistent one-loop calculation}
\label{sec:Self-consistent}

In this Section, we present a {\it self-consistent one-loop calculation}
of the equation for the density profile $\rho({\bf r})$
to second order in the gradient expansion.
The calculation involves taking the equations for the density in the 
Hartree-Fock approximation and expanding them around the Thomas-Fermi limit.
The result is the differential equation (\ref{rho-final})
that generalizes the Gross-Pitaevskii equation by taking into 
account the leading effects of quantum fluctuations. 

The self-consistent one-loop equations can be expressed as
classical field equations for the one-loop effective action 
\cite{Kirsten-Toms}.
We describe briefly the diagrammatic representation of these equations.
They correspond to summing all connected diagrams with arbitrarily
many one-loop subdiagrams, but no subdiagrams with two or more loops.
These diagrams have the structure of tree diagrams, 
with one-loop corrections added to the vertices 
and arbitrarily many one-loop corrections inserted 
into the propagators.  These diagrams can be calculated using the 
perturbative framework developed in Section \ref{sec:Cartesian}.
The sum of all such diagrams is independent 
of the choice of the background field $v({\bf r})$ in (\ref{psi-cart}).
However, the sum of all such diagrams can be greatly simplified
by choosing the background field $v$
so that the ground-state expectation values of the quantum fields
$\xi$ and $\eta$ vanish.  
This choice eliminates all one-particle-reducible diagrams which
can be disconnected by cutting a single $\xi$ or $\eta$ line.
The only diagrams that remain are one-particle-irreducible diagrams.

With the Cartesian parameterization (\ref{psi-cart}),
the choice of the background field that simplifies
self-consistent one-loop calculations is the condensate itself:
\begin{equation}
v({\bf r}) \;=\; \langle \psi({\bf r}) \rangle \,.
\label{v-cart}
\end{equation}
With this choice, the fields $\xi$ and $\eta$ represent the quantum 
fluctuations around the ground-state expectation  value of $\psi$.
Since $v$ is real-valued, the expectation value of $\eta$ vanished 
automatically and the condition (\ref{v-cart}) can be written
\begin{equation}
\langle \xi({\bf r}) \rangle  \;=\; 0 \,.
\label{tadpole-cart}
\end{equation}
Thus the background field $v$ must be chosen self-consistently so that the 
quantum fluctuations around that background average to zero. 
We will refer to the equation (\ref{tadpole-cart})
as the {\it tadpole equation}, because
the one-loop quantum corrections to this equation 
correspond to Feynman diagrams like those in Figure 6 and 7
that look like tadpoles. Using the tadpole equation,
the density (\ref{rho-cart}) reduces to 
\begin{equation}
\rho({\bf r}) \;=\;  v^2({\bf r})
\;+\; {1 \over 2} \left\langle \xi^2({\bf r}) \right\rangle
\;+\; {1 \over 2} \left\langle \eta^2({\bf r}) \right\rangle
\;-\; \delta \rho \,.
\label{rho-cart:0}
\end{equation}

The ground-state expectation values in (\ref{tadpole-cart}) 
and (\ref{rho-cart:0}) are nonlocal functionals of the background 
$v({\bf r})$.  Our strategy is to use the gradient expansion to 
reduce these functionals to local functions involving
$v({\bf r})$ and its derivatives.  The tadpole equation 
(\ref{tadpole-cart}) then 
reduces to an algebraic relation between $v({\bf r})$ and its derivatives,
while (\ref{rho-cart:0}) expresses $\rho$ in terms of 
$v$ and its derivatives.  If we eliminate $v$
from these two equations, we obtain an algebraic relation between 
$\rho$ and its derivatives.  This is the differential equation 
for $\rho({\bf r})$ that includes self-consistent corrections from 
one-loop quantum fluctuations.

To calculate the one-loop quantum corrections,
we use the decomposition (\ref{S-decomp}) of the action 
for quantum fluctuations around a general background field $v$.
The free part (\ref{S-free}) involves only the quantum fields
$\xi$ and $\eta$, but introduces an arbitrary scale $\Lambda$.
The interaction part (\ref{Sint-cart}) involves sources
$T$, $X$, $Y$, and $Z$ that are given in (\ref{T-def})--(\ref{Z-def}).
At one-loop order, the tadpole equation states that the expression 
(\ref{xi-m}) for $\langle \xi \rangle$ vanishes, which implies
\begin{equation}
0 \;=\;  T({\bf r}) 
\;+\; {3 \over 2}  Z({\bf r}) 
	\left\langle \xi^2({\bf r}) \right\rangle_{X,Y}
\;+\; {1 \over 2}  Z({\bf r}) 
	\left\langle \eta^2({\bf r}) \right\rangle_{X,Y} \,.
\label{tadpole-cart:1}
\end{equation}
Similarly, the expression (\ref{rho-cart:0}) for the number density 
reduces at one-loop order to 
\begin{equation}
\rho({\bf r}) \;=\; v^2({\bf r})
\;+\; {1 \over 2} \left\langle \xi^2({\bf r}) \right\rangle_{X,Y}  
\;+\; {1 \over 2} \left\langle \eta^2({\bf r}) \right\rangle_{X,Y}  
\;-\; \delta \rho \,.
\label{rho-cart:1}
\end{equation}
The equations (\ref{tadpole-cart:1}) and (\ref{rho-cart:1}) 
are integral equations whose solutions give the condensate 
and the density in the Hartree-Fock approximation.
The quantities $\langle \xi^2 \rangle_{X,Y}$
and $\langle \eta^2 \rangle_{X,Y}$ in (\ref{tadpole-cart:1}) 
and (\ref{rho-cart:1}) can be expanded in powers of $X$ and $Y$
and their derivatives using (\ref{xi-ms:DE}) and (\ref{eta-ms:DE}).
Since these expansions include infinitely many terms, the equations
(\ref{tadpole-cart:1}) and (\ref{rho-cart:1}) can be reduced to local 
equations only by consistently truncating the expansions.
We will reduce them to local equations at a specific point ${\bf r}_0$ by
(a) using the classical equations for $\rho$ and $v$ to simplify
the expressions for the sources, 
(b) choosing a specific value for the arbitrary parameter $\Lambda$,
and (c) truncating the equations at second order in the 
gradient expansion.

We begin by simplifying the sources $X$, $Y$, and $Z$
in (\ref{X-def}), (\ref{Y-def}), and (\ref{Z-def})
by using the classical equations 
$T({\bf r}) = 0$  and $\rho({\bf r}) = v^2({\bf r})$.
Since $X$, $Y$, and $Z$ appear only in one-loop diagrams,
any quantum corrections to the sources contribute only at 
second order in the quantum loop expansion.
Using $T = 0$, we can eliminate the potential $V$
from $X$ and $Y$.  Using $v = \sqrt{\rho}$, we can express 
$X$, $Y$, and $Z$ in terms of $\rho$ only.
We can also simplify $T$ by setting $v = \sqrt{\rho}$ in the terms 
proportional to the counterterms $\delta \mu$ and $\delta g$.
Finally, we can drop the terms in $X$ and $Z$
that involve the counterterm $\delta g$, since it is needed only
to cancel ultraviolet divergences that arise at two loops or higher 
in the quantum loop expansion.
Thus, the expressions for  the sources can be reduced to
\begin{eqnarray}
T({\bf r}) &=& 
\left[ \mu - V({\bf r}) 
 	- {g  \over 2} v^2({\bf r}) \right] v({\bf r})
\;+\; {1 \over 2 m} \mbox{\boldmath $\nabla$}^2 v({\bf r})
\;+\; \left[ \delta \mu - {\delta g \over 2} \rho({\bf r}) \right]
 	\sqrt{\rho}({\bf r}) \,,
\label{T-cl}
\\
X({\bf r}) &=& 
\Lambda^2 \;-\; 2 m g \rho({\bf r})
\;-\; {\mbox{\boldmath $\nabla$}^2 \sqrt{\rho} \over \sqrt{\rho}}({\bf r}) \,,
\label{X-cl}
\\
Y({\bf r}) &=&
- {\mbox{\boldmath $\nabla$}^2 \sqrt{\rho} \over \sqrt{\rho}}({\bf r}) \,,
\label{Y-cl}
\\
Z({\bf r}) &=& - {g \over 2} \sqrt{\rho}({\bf r}) \,.
\label{Z-cl}
\end{eqnarray}
Note that the source $Y$ in (\ref{Y-cl}) is already second order in the 
gradient expansion.  Thus if we truncate
the equations at second order  in the gradient expansion,
we need only include terms up to first order in $Y$
and we can omit all derivatives of $Y$.
We also need only include terms up to first order in 
$\mbox{\boldmath $\nabla$}^2 X$ and up to second order in
$\mbox{\boldmath $\nabla$} X$.  However, we still must include 
all possible powers of $X$.

In order to reduce the expansions for (\ref{tadpole-cart:1}) and
(\ref{rho-cart:1}) to a finite number of terms, we choose $\Lambda$ so that 
$X({\bf r}_0)$ is second order in the gradient expansion at a specific 
point ${\bf r}_0$.  
A convenient choice for $\Lambda$ is the wavenumber that appears in the 
Bogoliubov dispersion relation (\ref{epsilon}) for a homogeneous 
Bose gas with number density $\rho({\bf r}_0)$:
\begin{equation}
\Lambda^2 \;=\; 2 m g \rho({\bf r}_0) \,.
\label{Lambda-cart}
\end{equation}
With this choice for $\Lambda$, the source $X$ and its 
derivatives at the point ${\bf r}_0$ reduce to 
\begin{eqnarray}
X({\bf r}_0) &=& - 
{\mbox{\boldmath $\nabla$}^2 \sqrt{\rho} \over \sqrt{\rho}}({\bf r_0}) \,,
\label{X-r0}
\\
\mbox{\boldmath $\nabla$} X({\bf r}_0) &=& 	
- 4 m g \, \sqrt{\rho} \, \mbox{\boldmath $\nabla$} \sqrt{\rho}({\bf r}_0) \,,
\label{dX-r0}
\\
\mbox{\boldmath $\nabla$}^2 X({\bf r}_0) &=& 
- 4 m g \left[ \sqrt{\rho} \, \mbox{\boldmath $\nabla$}^2 \sqrt{\rho}
	+ (\mbox{\boldmath $\nabla$} \sqrt{\rho})^2 \right]({\bf r}_0) \,.
\label{ddX-r0}
\end{eqnarray}

We proceed to determine the differential equation satisfied by 
$\rho$ in the self-consistent one-loop approximation.  
This equation can be obtained by solving (\ref{rho-cart:1})
for the condensate $v$ in terms of the density $\rho$ and its derivatives,
and then eliminating $v$ from the tadpole equation (\ref{tadpole-cart:1}).
If the tadpole equation is evaluated at 
the point ${\bf r}_0$ and then truncated at second order in the 
gradient expansion, it reduces to 
\begin{eqnarray}
0 \;=\; T({\bf r}_0)
&+& Z({\bf r}_0) \left[ 
{3 a_0 + b_0 \over 2} \;+\; {3 a_1 + b_1 \over 2} \, X({\bf r}_0) 
	\;+\; {3 a_2 + b_2 \over 2} \, Y({\bf r}_0)
\right.
\nonumber \\
&& \left. \qquad
\;+\; {3 a_3 + b_3 \over 2} \mbox{\boldmath $\nabla$}^2 X({\bf r}_0) 
\;+\; {3 a_5 + b_5 \over 2} 
	( \mbox{\boldmath $\nabla$} X )^2({\bf r}_0)  \right] \,.
\label{tadpole-cart:2}
\end{eqnarray}
The expression (\ref{T-cl}) for the source $T$ involves $v$ and 
$\mbox{\boldmath $\nabla$}^2 v$.  Solving (\ref{rho-cart:1}) 
for $v({\bf r})$ to first order in the quantum corrections, we obtain
\begin{eqnarray}
v({\bf r}) &=& \sqrt{\rho}({\bf r})
\;-\; {1 \over 2 \sqrt{\rho}({\bf r})}
\left[ {a_0 + b_0 - 2 \delta \rho \over 2}
	\;+\; {a_1+ b_1 \over 2} \, X({\bf r}) 
	\;+\; {a_2 + b_2 \over 2} \, Y({\bf r})
\right.
\nonumber \\
&& 
\left. 
\;+\; {a_3 + b_3 \over 2} \mbox{\boldmath $\nabla$}^2 X({\bf r}) 
\;+\; {a_4 + b_4 \over 2} \, X^2({\bf r})
\;+\; {a_5 + b_5 \over 2} 
	( \mbox{\boldmath $\nabla$} X )^2({\bf r}) 
\;+\; \ldots \right]  \,.
\label{v-cart:1}
\end{eqnarray}
We then substitute this expression for $v$ into the 
source $T({\bf r})$ in (\ref{T-cl}) and expand to first order
in quantum fluctuations.
After calculating the derivative $\mbox{\boldmath $\nabla$}^2 v$
appearing in $T$, we can set ${\bf r} = {\bf r}_0$ 
and then truncate at second order in the gradient expansion.
Using the expressions (\ref{Y-cl}) and (\ref{X-r0})-(\ref{ddX-r0}) 
for the sources and their derivatives, 
the expression for $T({\bf r}_0)$ reduces to
\begin{eqnarray}
T &=&
\left( \mu - V - {g \over 2} \rho \right) \sqrt{\rho}
\;+\; {1 \over 2m} \mbox{\boldmath $\nabla$}^2 \sqrt{\rho}
\;+\; \left( {a_0 + b_0 - 2 \delta \rho \over 4} + {\delta \mu \over g} \right) 
	g \sqrt{\rho}
\;-\; {\delta g \over 2} \rho \sqrt{\rho}
\nonumber \\
&& 
\;+\; \left( {a_0 + b_0 - 2 \delta \rho \over 4} 
	+ {a_1 + b_1 - a_2 - b_2 \over 8} \Lambda^2
	- {a_3 + b_3 \over 4} \Lambda^4 \right)
	{\mbox{\boldmath $\nabla$}^2 \sqrt{\rho} \over m \rho}
\nonumber \\
&& 
\;-\; \left( {a_0 + b_0 - 2 \delta \rho \over 4} 
	+ {a_1 + b_1 \over 4} \Lambda^2
	+ {a_3 + b_3 + 4 a_4 + 4 b_4 \over 4} \Lambda^4 
	- {a_5 + b_5 \over 2} \Lambda^6 \right)
	{(\mbox{\boldmath $\nabla$} \sqrt{\rho})^2 \over m \rho \sqrt{\rho}}
\nonumber \\
&& \;+\; \left( v - \sqrt{\rho} \right)
\left[  \mu - V - {g \over 2} \rho 
+ {1 \over 2 m} {\mbox{\boldmath $\nabla$}^2 \sqrt{\rho} \over \sqrt{\rho}}
	\right] \,,
\label{Trho-cart}
\end{eqnarray}
where we have used the expression (\ref{Lambda-cart}) for $\Lambda$.
The last term in (\ref{Trho-cart}) can be dropped, because it is
proportional to the classical equation (\ref{GP-rho}).  Its effects are 
therefore of  second order in the quantum loop expansion.
This eliminates 
all occurrences of the potential $V$ in the quantum corrections.
Inserting the resulting expression for $T({\bf r}_0)$  into
(\ref{tadpole-cart:2}),  the tadpole equation reduces to
\begin{eqnarray}
0 &=& \left( \mu - V - {g \over 2} \rho \right) \sqrt{\rho}
\;+\; {1 \over 2 m} \mbox{\boldmath $\nabla$}^2 \sqrt{\rho}
\;-\; \left( {a_0 + \delta \rho \over 2} - {\delta \mu \over g} \right) 
	g \sqrt{\rho}
\;-\; {\delta g \over 2}  \rho \sqrt{\rho}
\nonumber \\
&& 
\;+\; \left( {a_0 + b_0 - 2 \delta \rho \over 4} 
	+ {2 a_1 + b_1 + a_2 \over 4} \Lambda^2
	+ {a_3 \over 2} \Lambda^4 \right)
	{\mbox{\boldmath $\nabla$}^2 \sqrt{\rho} \over m \rho}
\nonumber \\
&& 
\;-\; \left( {a_0 + b_0 - 2 \delta \rho \over 4} 
	+ {a_1 + b_1 \over 4} \Lambda^2 
	- {a_3 - 2 a_4 - 2 b_4 \over 2} \Lambda^4
	+ a_5 \Lambda^6 \right)
	{(\mbox{\boldmath $\nabla$} \sqrt{\rho})^2 \over m \rho \sqrt{\rho}} \,.
\label{tadpole-cart:3}
\end{eqnarray}
Using the results for the coefficients $a_i$ and $b_i$ given in
Appendix \ref{app:Coefficients} and using (\ref{Lambda-cart}) to set 
$\Lambda^2 = 2 m g \rho({\bf r}_0)$, the equation for $\rho$ reduces to
\begin{eqnarray}
0 &=& ( \mu - V({\bf r}) ) \sqrt{\rho}({\bf r}) 
\;-\; {g \over 2} \, \rho \sqrt{\rho}({\bf r})
\;+\; {1 \over 2 m} \mbox{\boldmath $\nabla$}^2 \sqrt{\rho}({\bf r})
\nonumber \\
&& 
\;-\; {1 \over 48 \pi^2} (2 m g)^{3/2} 
\left[ 4 g \, \rho^2({\bf r})
\;+\; {17 \over 24 m} 
\left( 2 \sqrt{\rho} \, \mbox{\boldmath $\nabla$}^2 \sqrt{\rho}({\bf r}) 
	+ (\mbox{\boldmath $\nabla$} \sqrt{\rho})^2({\bf r}) \right) 
\right] \,.
\label{tadpole-cart:5}
\end{eqnarray}
We derived this equation at the point ${\bf r}_0$ defined by
our choice (\ref{Lambda-cart}) for the arbitrary parameter $\Lambda$.
However, our final result is an algebraic equation 
relating $\sqrt{\rho}$ and its 
derivatives at the point ${\bf r}_0$.  Since we could have chosen
any specific point for ${\bf r}_0$, this algebraic relation must 
hold at any point ${\bf r}$.  
Using (\ref{g-a}) to eliminate $g$ in favor of $a$ and
using dimensional analysis to insert the appropriate factors of $\hbar$
into our equation (\ref{tadpole-cart:5}),
we obtain the differential equation (\ref{rho-final}) for the density profile.

The $\rho^2$ term in (\ref{tadpole-cart:5}) can be obtained from 
previous work on the homogeneous Bose gas. 
Differentiating the result (\ref{E-Yang})
for the energy density with respect to $\sqrt{\rho}$, we obtain
\begin{equation}
{\partial {\cal E} \over \partial \sqrt{\rho}}
\;=\;  g \rho \sqrt{\rho} 
\left[ 1 \;+\; {1 \over 6 \pi^2} (2 m g)^{3/2} \sqrt{\rho} \right]  \,.
\label{dE-Yang}
\end{equation}
Multiplying by $-{1 \over 2}$, we reproduce the $\rho \sqrt{\rho}$ and
$\rho^2$ terms in (\ref{tadpole-cart:5}).  
The $\sqrt{\rho} \, \mbox{\boldmath $\nabla$}^2 \sqrt{\rho}$
and $(\mbox{\boldmath $\nabla$} \sqrt{\rho})^2$ terms in 
(\ref{tadpole-cart:5}) are new results.

As a check of the equation (\ref{tadpole-cart:5}),
we can verify that our one-loop expression for $\rho$ given in 
(\ref{rho-1loop:2}) satisfies (\ref{tadpole-cart:5}) after expanding 
to first order in the quantum fluctuations.
There is an important qualitative difference between the 
approximate solution (\ref{rho-1loop:2}) and the solution to the 
self-consistent equation (\ref{tadpole-cart:5}).  The solution to 
(\ref{tadpole-cart:5}) has the correct qualitative behavior 
even outside the condensate.
In this region, the density is very small  and only the terms in 
(\ref{tadpole-cart:5}) that are linear in $\sqrt{\rho}$ are important.
The equation therefore reduces to 
\begin{equation}
0 \;\approx\; ( \mu - V({\bf r}) ) \sqrt{\rho}({\bf r}) 
\;+\; {1 \over 2 m} \mbox{\boldmath $\nabla$}^2 \sqrt{\rho}({\bf r}) \,.
\label{tadpole-asymptotic}
\end{equation}
The quantum correction terms in (\ref{tadpole-cart:5}) 
were calculated using a gradient
expansion that is valid only inside the condensate. 
However, since these terms are all higher order in $\sqrt{\rho}$, 
their effects are negligible outside the condensate and it does no
harm to include them.  In contrast, the approximate solution
(\ref{rho-1loop:2}) has the wrong qualitative behavior when 
$\phi_0$ is small, because it is dominated by the 
$\mbox{\boldmath $\nabla$}^2 \phi_0$ and 
$(\mbox{\boldmath $\nabla$} \phi_0)^2/\phi_0$ terms.
Thus that solution can only be used inside the condensate.

\section{Polar field parameterization}
\label{sec:Polar}

The one-loop calculations in Sections \ref{sec:One-loop} and 
\ref{sec:Self-consistent} were carried out using the 
Cartesian parameterization of the quantum field given in (\ref{psi-cart}).
There is nothing special about this parameterization
aside from its simplicity.  
Other field parameterizations should give
the same final result for physical quantities.
An example of an alternative field parameterization
is the {\it polar parameterization}:
\begin{equation}
\psi({\bf r},t) \;=\;  
\sqrt{v^2({\bf r}) + \sigma({\bf r},t)} \; \exp ( i \alpha({\bf r},t) ) \,.
\label{psi-polar}
\end{equation}
The advantage of this parameterization is that it eliminates
infrared divergences from individual Feynman diagrams that contribute 
to the number density.  In this Section, we verify that the polar 
parameterization gives the same equation for the density profile.
We also use this parameterization to show that the gradient expansion 
of the density breaks down at fourth order.

With the polar parameterization (\ref{psi-polar}),
the choice for the background field $v$ that simplifies
self-consistent one-loop quantum corrections is the one specified by the  
tadpole equation
\begin{equation}
\langle \sigma({\bf r}) \rangle  \;=\; 0 \,.
\label{tadpole-polar}
\end{equation}
The expression (\ref{rho-QFT}) for the number density reduces to 
\begin{equation}
\rho({\bf r}) \;=\; v^2({\bf r}) \;-\; \delta \rho \,.
\label{rho-polar}
\end{equation}
Thus the choice of the background $v({\bf r})$ implied by the 
tadpole condition (\ref{tadpole-polar}) is 
\begin{equation}
v({\bf r})  \;=\; \sqrt{\rho({\bf r})} 
\;+\; {\delta \rho \over 2} {1 \over \sqrt{\rho({\bf r})}} 
\;+\; \ldots \,.
\label{v-polar:2}
\end{equation}
Since this expression involves the ultraviolet divergent constant 
$\delta \rho$, $v$ has no simple physical interpretation.
It is best regarded as a theoretical construct 
that should appear only in intermediate stages of a calculation.
The simplicity of the expression (\ref{rho-polar}) for $\rho$
comes at the expense of the expression for 
the condensate.  Expanding (\ref{psi-polar}) as a power series 
in $\sigma$ and $\alpha$ and taking the ground-state expectation value, 
we obtain
\begin{equation}
\langle \psi({\bf r}) \rangle \;=\; v({\bf r}) 
\;-\; { 1 \over 8 v^3({\bf r})} \langle \sigma^2({\bf r}) \rangle 
\;-\; { v({\bf r}) \over 2} \langle \alpha^2({\bf r}) \rangle 
\;+\; \ldots \,.
\label{vev-polar}
\end{equation}
The expansion (\ref{vev-polar}) includes infinitely many terms and we
have written explicitly only those terms that contribute at one-loop order.
The expectation values of operators involving four or more powers of
$\sigma$ or $\alpha$ contribute at two-loop order or higher.

We begin our calculation by inserting the parameterization (\ref{psi-polar})
into the action (\ref{action}) and expanding in powers of the quantum fields
$\sigma$ and $\alpha$:
\begin{eqnarray}
S[\psi] \;=\;   S[v] &+& \int dt \int d^3r \Bigg\{ 
{1 \over v} T \sigma
\;+\; {1 \over 2} \left( \alpha \dot{\sigma} - \sigma \dot{\alpha} \right)
\;-\; {v^2 \over 2 m} (\mbox{\boldmath $\nabla$} \alpha)^2
\nonumber \\
&-& {1 \over 8 m v^2} (\mbox{\boldmath $\nabla$} \sigma)^2
\;-\; \left( {g + \delta g\over 4} 
	+ {\mbox{\boldmath $\nabla$}^2 v \over 4 m v^3}
	- {(\mbox{\boldmath $\nabla$} v)^2 \over 4 m v^4} \right) \sigma^2
\nonumber \\
&-& {1 \over 2m} \sigma (\mbox{\boldmath $\nabla$} \alpha)^2
\;+\; {1 \over 8 m v^4} \sigma (\mbox{\boldmath $\nabla$} \sigma)^2 
\;+\;  \left( {\mbox{\boldmath $\nabla$}^2 v \over 6 m v^5}
	- {(\mbox{\boldmath $\nabla$} v)^2 \over 3 m v^6} \right) \sigma^3
\,+\,  \ldots  
\Bigg\} ,
\label{S-polar}
\end{eqnarray}
where $T$ is the external source given in (\ref{T-def}).
The parameterization (\ref{psi-polar}) leads to 
an infinite series of momentum-dependent interactions.
We have dropped terms that are 
fourth and higher order in the quantum fields, 
since they do not contribute to the one-loop quantum corrections to the  
density profile.  It is convenient to introduce an arbitrary parameter 
$\Lambda$ into the action by rescaling the quantum fields as follows:
\begin{eqnarray}
\sigma({\bf r},t) \;=\; {\Lambda \over \sqrt{m g}} \, \xi({\bf r},t) \; ,
\\
\alpha({\bf r},t) \;=\; {\sqrt{m g} \over \Lambda} \, \eta({\bf r},t) \; .
\end{eqnarray}
After these rescalings, we separate the action into a free part and an 
interaction part as in (\ref{S-decomp}).  
The free part is identical to (\ref{S-free}) and 
the interaction part is
\begin{eqnarray}
S_{\rm int}[v,\xi,\eta] &=&   \int dt \int d^3r \Bigg\{ 
\sqrt{2} {\Lambda \over \sqrt{2 m g v^2}} T \xi
\;+\; {1 \over 4m} X \xi^2
\;+\; {1 \over 4m} U (\mbox{\boldmath $\nabla$} \xi)^2
\;+\; {1 \over 4m} S \left( \mbox{\boldmath $\nabla$} \eta \right)^2
\nonumber \\
&& 
\;+\; {1 \over \sqrt{2}} Z \xi^3
\;+\; {1 \over \sqrt{2}} W \xi (\mbox{\boldmath $\nabla$} \xi)^2
\;-\; {1 \over \sqrt{8} m v} {\sqrt{2 m g v^2} \over \Lambda}  
	\xi (\mbox{\boldmath $\nabla$} \eta)^2
\;+\;  \ldots  
\Bigg\}\,,
\label{Sint-polar}
\end{eqnarray}
where $X$, $Z$, $U$, $S$, and $W$ are external sources that 
depend on the background $v$:
\begin{eqnarray}
X({\bf r}) &=& - 2 \, {\Lambda^2 \over 2 m g v^2({\bf r})} 
	\left[ {\mbox{\boldmath $\nabla$}^2 v \over v}
	- {(\mbox{\boldmath $\nabla$} v)^2 \over v^2} \right] ({\bf r}) 
	- {\delta g \over g} \Lambda^2 \,,
\label{X'-def}
\\
Z({\bf r}) &=& {2 \over 3 m v({\bf r})} 
	\left( {\Lambda^2 \over 2 m g v^2({\bf r})} \right)^{3/2} 
	\left[ {\mbox{\boldmath $\nabla$}^2 v \over v}
	- 2 {(\mbox{\boldmath $\nabla$} v)^2 \over v^2} \right] ({\bf r}) \,,
\label{Z'-def}
\\
U({\bf r}) &=& 1 \;-\; {\Lambda^2 \over 2 m g v^2({\bf r})} \,,
\label{U-def}
\\
S({\bf r}) &=& 1 \;-\; {2 m g v^2({\bf r}) \over \Lambda^2} \,,
\label{V-def}
\\
W({\bf r}) &=& {1 \over 2 m v({\bf r})} 
	\left( {\Lambda^2 \over 2 m g v^2({\bf r})} \right)^{3/2} \,,
\label{W-def}
\end{eqnarray}
The arbitrary parameter $\Lambda$, which was introduced through
the rescaling of the fields, appears in both the free part 
of the action and the interactions.
We will exploit the arbitrariness of this
parameter to simplify the calculation of quantum corrections.

The tadpole equation (\ref{tadpole-polar}) can be written
$\langle \xi({\bf r}) \rangle  = 0$.  To first order in the
quantum corrections, this equation implies that
$(\Lambda/\sqrt{2 m g v^2}) T$ plus the sum of all one-loop tadpole
diagrams vanishes.  The one-loop diagrams include all possible 
insertions of the sources $X$, $U$, and $S$.
The one-loop tadpole equation can be written succinctly in the form
\begin{eqnarray}
0 \;=\; {\Lambda \over \sqrt{2 m g v^2}} \,  T 
&+& {3 \over 2} Z \, \left\langle \xi^2 \right\rangle_{X,U,S}
\;-\; {1 \over 2} \mbox{\boldmath $\nabla$} \cdot 
	\left[ W \mbox{\boldmath $\nabla$} 
	\left\langle \xi^2 \right\rangle_{X,U,S} \right]
\nonumber \\
&+& {1 \over 2} W \, 
	\left\langle (\mbox{\boldmath $\nabla$} \xi)^2 \right\rangle_{X,U,S}
\;-\; {1 \over 4 m v} {\sqrt{2 m g v^2} \over \Lambda}
	\left\langle (\mbox{\boldmath $\nabla$} \eta)^2 \right\rangle_{X,U,S} 
\,.
\label{tadpole-polar:2}
\end{eqnarray}
where $\langle \ldots \rangle_{X,U,S}$ denotes the ground-state 
expectation value in the presence of the sources $X$, $U$, and $S$, 
but with no other interactions.  The expectation values in 
(\ref{tadpole-polar:2}) are nonlocal functionals of these sources.
After Taylor-expanding the sources around the point ${\bf r}$,
these functionals can be expanded in powers of 
$X$, $U$, and $S$ and their derivatives at the point ${\bf r}$:
\begin{eqnarray}
\left\langle \xi^2 \right\rangle_{X,U,S} 
&=& c_0 + c_1 U 
	+ c_2  S 
	+ c_3 \, U^2 
	+ c_4 \, U S 
	+ c_5 \, S^2 \;+\; \ldots \,,
\label{xixi-m}
\\
\left\langle (\mbox{\boldmath $\nabla$} \xi)^2 \right\rangle_{X,U,S} 
&=& d_0 + d_1 \, X 
	+ d_2 \, \mbox{\boldmath $\nabla$}^2 U 
	+ d_3 \, \mbox{\boldmath $\nabla$}^2 S
\nonumber \\
&& \;+\; d_4 \left( \mbox{\boldmath $\nabla$} U \right)^2 
	+ d_5 \, \mbox{\boldmath $\nabla$} U \cdot 
		\mbox{\boldmath $\nabla$} S
	+ d_6  \left( \mbox{\boldmath $\nabla$} S \right)^2
	+ \ldots \,,
\label{dxi-ms}
\\
\left\langle (\mbox{\boldmath $\nabla$} \eta)^2 \right\rangle_{X,U,S} 
&=& e_0 + e_1 \, X
	+ e_2 \, \mbox{\boldmath $\nabla$}^2 U 
	+ e_3 \, \mbox{\boldmath $\nabla$}^2 S 
\nonumber \\
&& \;+\; e_4 \left( \mbox{\boldmath $\nabla$} U \right)^2 
	+ e_5 \mbox{\boldmath $\nabla$} U \cdot \mbox{\boldmath $\nabla$} S
	+ e_6 \left(\mbox{\boldmath $\nabla$} S \right)^2
	+ \ldots \,.
\label{deta-ms}
\end{eqnarray}
The terms on the right sides of (\ref{xixi-m})--(\ref{deta-ms}) 
include all combinations of $X$, $U$, $S$, and their derivatives 
that are allowed by 
rotational symmetry. We have written explicitly only those terms that 
are required to calculate the equation for the density through second
order in the gradient expansion.

The right side of the equation (\ref{tadpole-polar:2}) 
is a nonlocal functional of $v$.  When it is expanded in powers of 
$X$, $U$, $S$, and their derivatives, there are infinitely many terms.
The equation can be reduced to a local one
only by consistently truncating the expansions.
We will reduce (\ref{tadpole-polar:2}) to a local equation 
at a specific point ${\bf r}_0$ by
(a) choosing a specific value for the arbitrary parameter $\Lambda$,
and (b) truncating the equations at second order in the 
gradient expansion.
Since the sources $X$ and $Z$ in (\ref{X'-def}) and (\ref{Z'-def})
are already second order in the gradient expansion, we need only keep terms 
of first order in $X$ and $Z$ and we can omit any derivatives of $X$.  
Moreover, we need only include terms that are first order in 
$\mbox{\boldmath $\nabla$}^2 U$ and $\mbox{\boldmath $\nabla$}^2 S$
and terms up to second order in 
$\mbox{\boldmath $\nabla$} U$ and $\mbox{\boldmath $\nabla$} S$.
However the equation (\ref{tadpole-polar:2}) still includes 
all possible powers of $U$ and $S$.  
We can reduce this equation to a finite number of terms at
a specific point ${\bf r}_0$ by choosing $\Lambda$ so that
the sources $U$ and $S$ vanish at the point ${\bf r}_0$.
The required value is 
\begin{equation}
\Lambda^2 \;=\; 2 m g v^2({\bf r}_0) \,.
\label{Lambda-polar}
\end{equation}
With this choice of $\Lambda$, the sources on the right sides of
(\ref{xixi-m})--(\ref{deta-ms}) reduce to
\begin{eqnarray}
X({\bf r}_0) &=& - 2 
\left[ {\mbox{\boldmath $\nabla$}^2 v \over v}
	-  { (\mbox{\boldmath $\nabla$} v)^2 \over v^2 }
	\right] ({\bf r}_0) \,,
\label{X'-r0}
\\
U({\bf r}_0) 
&=& 0 \,,
\label{U-r0}
\\
\mbox{\boldmath $\nabla$} U({\bf r}_0) 
&=& 2 \, {\mbox{\boldmath $\nabla$} v \over  v}({\bf r}_0) \,,
\label{dU-r0}
\\
\mbox{\boldmath $\nabla$}^2 U({\bf r}_0) 
&=& 2
\left[ {\mbox{\boldmath $\nabla$}^2 v \over v}
	- 3 { (\mbox{\boldmath $\nabla$} v)^2 \over v^2} 
	\right] ({\bf r}_0) \,,
\label{ddU-r0}
\\
S({\bf r}_0) 
&=& 0 \,,
\label{V-r0}
\\
\mbox{\boldmath $\nabla$} S({\bf r}_0) 
&=& - 2 \, {\mbox{\boldmath $\nabla$} v \over  v}({\bf r}_0) \,,
\label{dV-r0}
\\
\mbox{\boldmath $\nabla$}^2 S({\bf r}_0) 
&=& - 2 
\left[ {\mbox{\boldmath $\nabla$}^2 v \over v}
	+ { (\mbox{\boldmath $\nabla$} v)^2 \over v^2}
	\right]({\bf r}_0) \,,
\label{ddV-r0}
\end{eqnarray}
In the expression (\ref{X'-r0}) for $X({\bf r}_0)$, 
we have dropped the term involving the counterterm $\delta g$,
since it is needed only
to cancel ultraviolet divergences that arise at two loops or higher.
With the choice (\ref{Lambda-polar}) for $\Lambda$,
the tadpole equation (\ref{tadpole-polar:2}) simplifies at the 
point ${\bf r}_0$ to
\begin{eqnarray}
0 \;=\;   T 
&+& {1 \over m} \left[ {\mbox{\boldmath $\nabla$}^2 v \over v^2}
		- 2 { (\mbox{\boldmath $\nabla$} v)^2 \over v^3} \right] 
	\left\langle \xi^2 \right\rangle_{X,U,S}
\;-\; {1 \over 4 m v} \mbox{\boldmath $\nabla$}^2  
	\left\langle \xi^2 \right\rangle_{X,U,S}
\nonumber \\
&+&
{\mbox{\boldmath $\nabla$} v \over m v^2} \cdot 
	\mbox{\boldmath $\nabla$} \left\langle \xi^2 \right\rangle_{X,U,S} 
\;+\; {1 \over 4 m v} 
\left( 
\left\langle (\mbox{\boldmath $\nabla$} \xi)^2 \right\rangle_{X,U,S}
- \left\langle (\mbox{\boldmath $\nabla$} \eta)^2 \right\rangle_{X,U,S} 
\right) \,.
\label{tadpole-polar:3}
\end{eqnarray}
After inserting the expansions (\ref{xixi-m})--(\ref{deta-ms}) 
into the tadpole equation (\ref{tadpole-polar:3})
and evaluating it at ${\bf r}_0$, we can truncate
it at second order in the gradient expansion.  Using the
expressions (\ref{X'-r0})-(\ref{ddV-r0}) for the sources,
we obtain an algebraic equation relating  
$v$, $\mbox{\boldmath $\nabla$} v$, and $\mbox{\boldmath $\nabla$}^2 v$
at the point ${\bf r}_0$.  
To express this equation in terms of $\rho$ and its derivatives,
we eliminate $v$ using (\ref{v-polar:2}).
Since the term $\delta \rho/(2 \sqrt{\rho})$ in (\ref{v-polar:2}) is 
first order in quantum fluctuations, it is needed 
only in the term $T$ in (\ref{tadpole-polar:3}).  
The source $T$ then becomes
\begin{eqnarray}
T &=&
\left( \mu - V - {g \over 2} \rho \right) \sqrt{\rho}
\;+\; {1 \over 2m} \mbox{\boldmath $\nabla$}^2 \sqrt{\rho}
\;-\; \left( {\delta \rho \over 2} - {\delta \mu \over g} \right) 
	g \sqrt{\rho}
\;-\; {\delta g \over 2} \rho \sqrt{\rho}
\nonumber \\
&& 
\;-\; {\delta \rho \over 2} \left(
	{\mbox{\boldmath $\nabla$}^2 \sqrt{\rho} \over m \rho}
\;-\; {(\mbox{\boldmath $\nabla$} \sqrt{\rho})^2 
	\over m  \rho \sqrt{\rho}} \right)
\;+\; {\delta \rho \over 2}
\left[ \left( \mu - V - {g \over 2} \rho \right) {1 \over \sqrt{\rho}} 
\;+\; {1 \over 2 m} {\mbox{\boldmath $\nabla$}^2 \sqrt{\rho} \over \rho}
	\right] \,.
\label{Trho-polar}
\end{eqnarray}
The last term in (\ref{Trho-polar}) can be dropped, because it is
proportional to the classical equation (\ref{GP-rho}).  Its effects are 
therefore second order in the quantum loop expansion.
After inserting (\ref{Trho-polar}) into (\ref{tadpole-polar:3}), 
the tadpole equation reduces to 
\begin{eqnarray}
0 &=& \left( \mu - V - {g \over 2} \rho \right) \sqrt{\rho}
\;+\; {1 \over 2m} \mbox{\boldmath $\nabla$}^2 \sqrt{\rho}
\nonumber \\
&& 
\;+\; {d_0 - e_0 \over 4} {1 \over m  \sqrt{\rho}}
\;+\; \left( {\delta \mu \over g} - {\delta \rho \over 2} \right) g \sqrt{\rho}
	\;-\; {\delta g \over 2}   \rho \sqrt{\rho}
\nonumber \\
&& 
\;+\; \left( c_0 - {c_1 - c_2 \over 2} - {d_1 - d_2 + d_3 \over 2} 
	+ {e_1 - e_2 + e_3 \over 2} - {\delta \rho \over 2} \right)
	{\mbox{\boldmath $\nabla$}^2 \sqrt{\rho} \over m \rho}
\nonumber \\
&& 
\;-\; \left(2 c_0 - {7c_1 - 3 c_2 \over 2} + 2 (c_3 - c_4 + c_5)
	- {d_1 - 3 d_2 - d_3  \over 2} - (d_4 - d_5 + d_6)
	\right.
\nonumber \\
&& \qquad \qquad
\left. + {e_1 - 3 e_2 - e_3 \over 2} + (e_4 - e_5 + e_6) 
	- {\delta \rho \over 2} \right)
	{(\mbox{\boldmath $\nabla$} \sqrt{\rho})^2 \over m  \rho \sqrt{\rho}}  
\,.
\label{rho-polar:2}
\end{eqnarray}
The coefficients $c_i$, $d_i$, and $e_i$ are given in  
Appendix \ref{app:Coefficients}.  
The coefficients are all infrared finite but ultraviolet divergent.  
The ultraviolet divergences from individual diagrams are more severe
than those encountered with the Cartesian field parameterization 
used in Sections \ref{sec:One-loop} and \ref{sec:Self-consistent}.
The integrals $d_0$ and $e_0$ diverge as the fifth power of the 
ultraviolet cutoff, but they cancel in the combination $d_0 - e_0$.
The remaining ultraviolet divergences are cancelled by the
counterterms $\delta \mu$, $\delta g$, and $\delta \rho$, whose values 
are given in (\ref{delmu}), (\ref{delg}), and (\ref{delrho}). 

Using the expression for the coefficients given in the Appendix,
the equation (\ref{rho-polar:2}) reduces to
\begin{eqnarray}
0 &=& ( \mu - V ) \sqrt{\rho}
\;-\; {g \over 2} \rho \sqrt{\rho}
\;+\; {1 \over 2 m} \mbox{\boldmath $\nabla$}^2 \sqrt{\rho}
\nonumber \\
&& 
\;-\; { 1 \over 48 \pi^2} 
\left[ 2 {\Lambda^5 \over m \sqrt{\rho}}
\;+\; {17 \over 12} {\Lambda^3  \over m \rho}
	\mbox{\boldmath $\nabla$}^2 \sqrt{\rho}
\;+\; {17 \over 24} {\Lambda^3 \over m \rho \sqrt{\rho}}
	(\mbox{\boldmath $\nabla$} \sqrt{\rho})^2 \right] \;.
\label{rho-polar:3}
\end{eqnarray}
After using (\ref{Lambda-polar}) to set $\Lambda^2 = 2 m g \rho({\bf r}_0)$,
we reproduce the self-consistent one-loop 
equation (\ref{tadpole-cart:5}) for the density profile.

There has been a previous attempt to calculate the quantum corrections 
to the Gross-Pitaevskii equation \cite{Ilinski-Stepanenko}.
The authors used the polar field parameterization (\ref{psi-polar}),
with the background field $v({\bf r})$ equal to the 
mean field $\phi_0({\bf r})$.
They dropped all terms in the action that were third order and higher 
in $\alpha$ and $\sigma$, and they also dropped second order terms that
involved gradients of $\phi_0$ or $\sigma$.  The only terms remaining 
in the action that contribute to the density profile are
\begin{equation}
S[\psi] \;=\;   S[\phi_0] 
\;+\; \int dt \int d^3r \left\{ 
{1 \over 2} \left( \alpha \dot{\sigma} - \sigma \dot{\alpha} \right)
\;-\; {\phi_0^2({\bf r}) \over 2 m} (\mbox{\boldmath $\nabla$} \alpha)^2
\;-\; {g \over 4} \sigma^2 \right\} \,.
\label{action:I-S}
\end{equation}
With such a drastic truncation of the action, the quantum corrections 
that they ultimately calculate are of no relevance to the problem 
of atoms in a trapping potential.  This is evident from the fact 
that the Bogoliubov dispersion relation (\ref{epsilon})
never enters into the quantum corrections that they calculate.  Thus,
their approach is incapable of reproducing the known
results for a homogeneous Bose gas.

A comparison of the calculation above with that presented in Section
\ref{sec:Self-consistent} 
demonstrates that the Cartesian field parametrization 
is more efficient than the polar field parameterization for explicit 
calculations.  With the polar field parameterization, 
one avoids infrared divergent 
integrals at intermediate stages of the calculation, 
but this advantage is compensated by the fact that the 
integrals are more severely ultraviolet divergent.
The simplicity of the relation (\ref{rho-polar}) between $\rho$ and $v$ 
is compensated by a tadpole equation (\ref{tadpole-polar:2})
that is more complicated than the corresponding equation
(\ref{tadpole-cart:1}) in the Cartesian field parameterization.  

The advantage of the polar field parameterization is that it avoids 
cancellations of infrared divergences between different diagrams.
This makes it easier to identify the sources of infrared divergences that 
are responsible for the breakdown of the 
gradient expansion.  We will use this parameterization to show that the 
gradient expansion of the density breaks down at fourth order.
The component of the propagator matrix (\ref{propagator})
that is most infrared sensitive is $D^{\eta \eta}$.
For small loop momentum $k$, the frequency $\omega$ in the loop 
scales like $\Lambda k/(2m)$ and $D^{\eta \eta}$ scales like
$2m/k^2$.  The most infrared singular diagrams are those 
for which all the lines are $\eta$ lines.  The term in the tadpole 
equation (\ref{tadpole-polar:2}) that is most infrared sensitive is 
$\langle (\mbox{\boldmath $\nabla$} \eta)^2 \rangle_{X,U,S}$,
because the operator 
$ (\mbox{\boldmath $\nabla$} \eta)^2$ creates two $\eta$ lines.
In the expansion (\ref{deta-ms}) for that matrix element,
the most infrared singular terms are those
that involve only the source $S$, which couples to
a pair of $\eta$ lines.  The infrared behavior of the 
coefficient of a term in (\ref{deta-ms}) that involves $m$ factors of 
$\mbox{\boldmath $\nabla$}$ 
and $n$ factors of $S$ can be determined by simple power counting.
The integrand has a factor of $1/k^2$ for each of the
$n+1$ propagators.  There is a factor of $k^2$ for the operator 
$(\mbox{\boldmath $\nabla$} \eta)^2$ and a factor of $k^2$ for 
each insertion of $S$. Finally, dimensional analysis requires that each factor 
of $\mbox{\boldmath $\nabla$}$ be compensated by a factor of $1/k$ 
in the integrand.   Thus the integral must scale like
\begin{equation}
\int d \omega \int d^3 k \;
	k^2 \left( {1 \over k^2} \right)^{n+1} (k^2)^n  
	\left( {1 \over k} \right)^m
\;\sim \; \int dk \, k^{3-m} \,.
\label{int-IR}
\end{equation}
An infrared divergence can appear only if $m \ge 4$.
Thus infrared divergences first appear in the tadpole equation
when it is expanded to fourth order in the gradient expansion.

Our explicit one-loop calculation of $\langle \psi \rangle$
in (\ref{vev-1loop:1}) showed that the gradient expansion for the
condensate  breaks down at second order.
The analysis presented above shows that the gradient expansion for the
density  $\rho$ does not break down until fourth order.
The breakdown of the gradient expansion implies that the 
quantum corrections depend on nonlocal effects involving the length scale
$L$ for significant variations in $\rho$. 
While we have identified
the orders at which the gradient expansions break down,
we have not identified any deep reason for the gradient expansion 
of the density to be better behaved than that of the condensate.

\section{Implications for Present Traps}
\label{sec:Imp}

In this Section, we estimate the magnitude of the effects of 
quantum fluctuations for Bose-Einstein condensates in existing 
magnetic traps.  For numerical estimates, we will use 
parameters characteristic of the sodium experiment \cite{MIT}, 
which has achieved the highest density condensates to date.  
In this experiment, $N \approx 5 \times 10^{6}$ sodium atoms 
were condensed in a trapping potential 
with a length scale $\ell \approx 2$ $\mu$m.
The S-wave scattering length for sodium atoms is $a \approx 0.005$ $\mu$m.
The number density that was attained at the center of the condensate was
$\rho \approx 400$/$\mu$m$^3$.

Baym and Pethick have presented  a simple qualitative 
analysis of the solution to the Gross-Pitaevskii equation 
that allows one to determine how various quantities scale
with the number $N$ of atoms \cite{Baym-Pethick}.  
The qualitative behavior of the solution to the Gross-Pitaevskii
equation depends crucially on 
a dimensionless parameter $\zeta$ given by 
\begin{equation}
\zeta \;=\; \left( {8 \pi N a \over \ell} \right)^{1/5} \,,
\label{zeta}
\end{equation}
where $\ell$ is the length scale associated with significant
variations in the potential $V({\bf r})$.
For a harmonic oscillator potential, $\ell= \hbar/\sqrt{m \omega}$, 
where $\omega$ is the angular frequency of the harmonic oscillator.
If $\zeta$ is less than or of order 1, the size of the condensate is
comparable to $\ell$ and the number density inside the 
trap scales like $N/\ell^3$. If $\zeta$ is much greater than 1, 
the size $L$ of the condensate scales like  $L = \zeta \ell$ and the 
density inside the trap scales like $N/(\zeta \ell)^3$.  
The condensates in existing magnetic traps are characterized by 
values of $\zeta$ that are significantly greater than 1.
(For the sodium experiment, $\zeta \approx 13$.)
We will determine how the correction terms 
in our equation for the density profile
scale with $N$ in the case  $\zeta \gg 1$.
The expansion parameter $\sqrt{\rho a^3}$ for the low-density expansion 
scales like $\zeta a/\ell$.  Although $\zeta$ is large for existing traps,
$a/\ell$ is tiny and the product $\zeta a/\ell$ is small.
(For the sodium experiment, $\zeta a/\ell \approx 0.03$.)
The modes that dominate the quantum corrections have wavelengths
on the order of $1/\sqrt{\rho a}$, which scales like $\ell/\zeta$.
Since this is small compared to the length scale $\zeta \ell$ 
for significant variations in $\rho({\bf r})$, it is reasonable
to expand the quantum corrections using the gradient expansion. 
The gradient expansion corresponds to an expansion in powers of 
the dimensionless quantity $1/(\sqrt{\rho a} \zeta \ell)$, 
which scales like $1/\zeta^2$.  Thus, inside the condensate, 
quantum corrections are suppressed by $\zeta a/\ell$ and
corrections from second order in the gradient expansion
are suppressed by $1/\zeta^4$.

Outside the condensate, the density $\rho$ rapidly approaches 0.
The only terms in the equation (\ref{rho-final}) that are important 
in this region are the terms that are linear in $\sqrt{\rho}$.
The scale of the gradient is now set by the length scale $\ell$ for
significant variations in $V({\bf r})$.  In this region, the 
basic assumption underlying our calculation, 
that the quantum corrections are dominated by
wavelengths of order $1/\sqrt{\rho a}$, breaks down completely.
However, all the quantum corrections terms are higher order 
in $\sqrt{\rho}$ and therefore have a negligible effect on the solution
outside the condensate.  Thus it does no harm to include the quantum
correction terms in (\ref{rho-final}) in the exterior region.

The crossover region between the interior and exterior of the condensate
can be characterized by the fact that the 
$\mbox{\boldmath $\nabla$}^2 \sqrt{\rho}$ term and the 
$\rho \sqrt{\rho}$ terms become comparable in importance.
The gradient expansion breaks down in this region.
At the beginning of the crossover region, 
$\mbox{\boldmath $\nabla$}$ still scales like $1/(\zeta \ell)$,
but the density has decreased to the point that $\rho$
scales like $a/(\zeta \ell)^2$.
Therefore the quantum loop expansion parameter $\sqrt{\rho a^3}$
scales like $a/(\zeta \ell)$.
As long as this quantity is sufficiently small,
the quantum correction terms in (\ref{rho-final}) are negligible.
(In the sodium experiment, we have $a/(\zeta \ell) \approx 0.0002$.)
Thus it does no harm to include the quantum
correction terms in (\ref{rho-final}) in the cross-over region.
We conclude that the differential equation (\ref{rho-final})
can be used to calculate the density profile everywhere. 

We now give a quantitative estimate of the error from truncating 
the quantum loop expansion at one-loop order.  
A simple estimate of the relative magnitude of 
the quantum corrections is the ratio of the $\rho^2$ correction term in 
(\ref{rho-final}) to the $\rho \sqrt{\rho}$
term, which is $(32/3) \sqrt{\rho a^3/\pi}$.  For the sodium experiment,
this ratio is approximately 0.04 at the  center of the condensate.
This is small enough that quantum corrections can be treated as 
small perturbation to the mean-field approximation.  Since 
$\sqrt{\rho a^3}$ scales like $N^{1/5}$, the number of atoms 
in the trap could be increased by many orders of magnitude
and the condensate would still be within the perturbative region.

A naive estimate of the relative magnitude of two-loop 
quantum corrections is the square of the magnitude 
of the one-loop quantum corrections.  
Their effects should therefore be negligible.
One complication is that the two-loop correction
depends not only on the S-wave scattering length $a$,
but also on a second parameter that represents 
a pointlike contribution to the $3 \to 3$ scattering amplitude 
\cite{Braaten-Nieto}.  If this parameter is anomalously large, 
the two-loop quantum corrections could be significantly larger 
than the naive estimate.

We next quantify the errors from truncating the equation for the 
density at second order in the gradient expansion.
A simple estimate of the relative magnitude of 
contributions from second order in the gradient expansion is the ratio of
the $\mbox{\boldmath $\nabla$}^2 \sqrt{\rho}$ term to the 
$\rho \sqrt{\rho}$ term in the Gross-Pitaevskii equation (\ref{GP-rho}).  
Assuming that $\mbox{\boldmath $\nabla$}^2 \sqrt{\rho}$
scales like $\sqrt{\rho}/(\zeta \ell)^2$, the ratio is
$1/(8 \rho a (\zeta \ell)^2)$.  In the sodium experiment,
this ratio is roughly $10^{-5}$ at the center of the condensate.  
These corrections are therefore negligible.  
The relative importance of quantum corrections that are second order 
in the gradient expansion increases as one approaches the 
edge of the condensate, where the gradient expansion breaks down.
However they are still suppressed by a quantum loop factor of
order $a/(\zeta \ell)$.
 
The gradient expansion for the density breaks down at fourth order,
but the breakdown in only logarithmic in $8 \rho a (\zeta \ell)^2$.
We can estimate the magnitude of these corrections 
by taking the logarithms to be of order 1.  These corrections are 
therefore suppressed by two powers of $1/(8 \rho a (\zeta \ell)^2)$.
There is also an additional suppression factor of $\sqrt{\rho a^3}$,
since terms of fourth order in the gradient expansion enter only 
through quantum corrections.  Thus these corrections should 
be completely negligible.

In this paper, we have developed a framework 
for calculating the dominant effects of quantum fluctuations in a 
Bose-Einstein condensate trapped by an external potential.
Our method is based on a combination of the Hartree-Fock approximation
and an expansion around the Thomas-Fermi limit.
We have illustrated the method by calculating the self-consistent 
one-loop equation for the density profile to second order in the 
gradient expansion and the relation between the condensate and the density to
zeroth order in the gradient expansion.
It should be straightforward to use this method to calculate the 
effects of quantum fluctuations
on other properties of the condensate at zero temperature, 
such as the spectrum of its collective excitations.
It would also be interesting to extend the method to nonzero temperature,
so that one could study how
the effects of quantum fluctuations vary with temperature.

This work was supported in part by the U.~S. Department of Energy,
Division of High Energy Physics, under Grant DE-FG02-91-ER40690.
We thank T.-L. Ho for valuable discussions.

\appendix
\renewcommand{\theequation}{\thesection\arabic{equation}}

\section{Integrals}
\label{app:Integrals}
\setcounter{equation}{0}

In this Appendix, we give analytic expressions for the frequency integrals 
and the wavevector integrals that are required to calculate the
one-loop quantum corrections to the density and the condensate.  

Since time-independent sources do not change the energy, the 
frequency integrals are rather simple.
They can be evaluated using contour integration.
The specific integrals that are required are
\begin{eqnarray}
\int{d \omega \over 2 \pi} {1 \over (\omega^2 - \epsilon^2)^n}
&=& i (-1)^{n+1} 
	{(-1) \cdot 1 \cdot 3  \cdots (2n-3) \over 2^n (n-1)!} 
	{1 \over \epsilon^{2n-1}} \,,
\label{int-omega}
\\ 
\int{d \omega \over 2 \pi} {\omega^2 \over (\omega^2 - \epsilon^2)^{n+1}}
&=& i (-1)^{n+1} 
	{(-1) \cdot 1 \cdot 3 \cdots (2n-3) \over 2^{n+1} n!} 
	{1 \over \epsilon^{2n-1}} \,.
\end{eqnarray}

Time-independent sources that are inserted into a loop diagram
change the wavevector ${\bf k}$ of the propagators in the loop.
The gradient expansion corresponds to 
expanding the loop integral in powers of the wavevectors
${\bf p}_i$ of the sources
and  of the external lines of the diagram. After averaging over 
integration angles, the $k$ integrals that are required 
have the form
\begin{equation}
I_{m,n} \;\equiv\; 
\int{d^3k \over (2 \pi)^3} {(k^2)^m \over (k \sqrt{k^2 + \Lambda^2})^n} \,,
\label{Imn-def}
\end{equation}
where $m$ and $n$ are integers.
If $m$ and $n$ satisfy $m-n < - {3 \over 2}$,
this integral is ultraviolet convergent.
If $2m-n > -3$, the integral is infrared convergent.
If it is both ultraviolet and infrared convergent, its value is 
\begin{equation}
I_{m,n} \;=\; 
{ \Gamma(n - m - {3 \over 2}) \Gamma({3-n \over 2} + m) 
	\over 4 \pi^2 \Gamma({n \over 2}) } \,
	\Lambda^{3+2m-2n}\,, 
\qquad m + {3 \over 2} < n < 2m+3 \,.
\label{Imn-conv}
\end{equation}

The ultraviolet-divergent integrals that are required are
$I_{m,n}$ for $n-m = -1,0,1$, which have power ultraviolet divergences.
A convenient way to regularize the 
integral is to subtract pure powers of $k$ from the 
integrand that will remove the ultraviolet divergence, 
and then to add those powers of $k$ back in with an ultraviolet 
cutoff $k < \Lambda_{\rm UV}$.  The regularized integral is then
\begin{eqnarray}
I_{m,n} &\equiv& 
{1 \over 2 \pi^2} \int_0^\infty dk 
\left( {k^{2+2m-n} \over (k^2 + \Lambda^2)^{n/2}} 
\;-\; \sum_{i=0}^{m-n+1} {- {n \over 2} \choose i} \Lambda^{2i} 
	k^{2+2m-2n-2i} \right)
\nonumber \\
&&
\;+\; {1 \over 2 \pi^2} 
	\sum_{i=0}^{m-n+1} {- {n \over 2} \choose i} \Lambda^{2i}
	\int_0^{\Lambda_{\rm UV}} dk \, k^{2+2m-2n-2i} \,,
\qquad n < m + {3 \over 2} \,.
\label{Imn-UV}
\end{eqnarray}
The first integral in (\ref{Imn-UV}) is convergent and 
is equal to the expression on the right side of  (\ref{Imn-conv}).
The only dependence on the ultraviolet cutoff comes from the
remaining integrals in (\ref{Imn-UV}), and they give a polynomial in  
$\Lambda_{\rm UV}$.
The ultraviolet divergent integrals that arise in our calculation are
\begin{eqnarray}
I_{n-1,n} &=& {1 \over 4 \pi^2} 
\left( 2 \, \Lambda_{\rm UV}
\;+\; {\Gamma(- {1 \over 2}) \Gamma({n+1 \over 2}) \over \Gamma({n \over 2})} 
	\, \Lambda \right) \,, 
\label{In-1n}
 \\
I_{n,n} &=& {1 \over 4 \pi^2} 
\left( {2 \over 3} \, \Lambda_{\rm UV}^3
\;-\; n \, \Lambda_{\rm UV} \Lambda^2 
\;+\; {\Gamma(-{3 \over 2}) \Gamma({n+3 \over 2}) \over \Gamma({n \over 2})} 
	\, \Lambda^3 \right) \,, 
\label{Inn}
 \\
I_{n+1,n} &=& {1 \over 4 \pi^2} 
\left(  {2 \over 5} \, \Lambda_{\rm UV}^5
\;-\; {n \over 3} \, \Lambda_{\rm UV}^3 \Lambda^2 
\;+\; {n (n+2) \over 4} \, \Lambda_{\rm UV} \Lambda^4 
\;+\; {\Gamma(- {5 \over 2}) \Gamma({n+5 \over 2}) \over \Gamma({n \over 2})} 
	\, \Lambda^5 \right) \,. 
\label{In+1n}
\end{eqnarray}

The infrared-divergent integrals that are required are
$I_{m,n}$ for $n = 2m+3$, which have logarithmic infrared divergences.
The integrals can be regularized by imposing an infrared
cutoff $k > \lambda_{\rm IR}$.  In the limit $\lambda_{\rm IR} \ll \Lambda$,
the regularized integral reduces to
\begin{equation}
I_{m,2m+3} \;\equiv\; 
{1 \over 2 \pi^2} \int_0^\infty dk 
\left( {1 \over k (k^2 + \Lambda^2)^{m+3/2}} 
\;-\; {\theta(\Lambda - k) \over k \Lambda^{2m+3}} \right)
\;+\; {1 \over 2 \pi^2} \Lambda^{-2m-3} 
	 \log {\Lambda \over \lambda_{\rm IR}} \,.
\label{Im2m+3}
\end{equation}
The specific integrals that are needed in our calculations are
\begin{eqnarray}
I_{-2,-1} &=& 
{1 \over 4 \pi^2} 
\left[ 2 \, \Lambda_{\rm UV} 
\;+\; \left( \log {4 \Lambda^2 \over \lambda_{\rm IR}^2} 
		\;-\; 2 \right) \Lambda 
\right] \,,
\label{I-2-1}
\\
I_{-1,1} &=& 
{1 \over 4 \pi^2} \left( \log {4 \Lambda^2 \over \lambda_{\rm IR}^2} \right)
	{1 \over \Lambda} \,,
\label{I-11}
\\
I_{0,3} &=& 
{1 \over 4 \pi^2} \left( \log {4\Lambda^2 \over \lambda_{\rm IR}^2} 
			\;-\; 2 \right)
	{1 \over \Lambda^3} \,.
\label{I03}
\end{eqnarray}
Note that the integral $I_{-2,-1}$ is ultraviolet divergent as 
well as infrared divergent.

\section{Explicit Calculation of a Diagram}
\label{app:Diagram}
\setcounter{equation}{0}

In this Appendix, we illustrate the calculation of one-loop diagrams 
that contribute to the quantum corrections to the density
by calculating one  diagram in detail.
We consider the last diagram in Figure 3, which represents a
contribution to the matrix element $\langle \xi^2 \rangle_{X,Y}$ 
involving two insertions of the source $X$.

It is convenient to calculate the diagram in wavevector space,
and then Fourier transform to get the diagram in coordinate space.
The diagram involves an integral 
over the frequency $\omega$ and an integral over the wavevector ${\bf k}$ 
running around the loop.  Letting the wavevectors 
of the two sources be ${\bf p}_1$ and ${\bf p}_2$,
the expression for the diagram is 
\begin{equation}
{1 \over 2} \int {d \omega \over 2 \pi}
\int{d^3k \over (2 \pi)^3} \; 2 \; 
{i ({\bf k} + {\bf p}_1)^2/(2 m) 
	\over \omega^2 - \epsilon^2(| {\bf k} + {\bf p}_1 |) }  \;
{i X({\bf p}_1) \over 2m}  \;
{i k^2/(2 m) \over \omega^2 - \epsilon^2(k)}  \;
{i X({\bf p}_2) \over 2m} \;
{i ({\bf k} - {\bf p}_2)^2/(2 m) 
	\over \omega^2 - \epsilon^2(| {\bf k} - {\bf p}_2 |) }  \,.
\end{equation}
We have written the Feynman rules for each of the propagators and vertices 
in the loop in the order in which they appear as you go clockwise 
around the loop.  There is a symmetry factor of ${1 \over 2}$ and
the factor of 2 inside the integral is the Feynman rule for the 
operator $\xi^2$. There is an implied  $+i  0^+$ prescription
in the denominator of each of the propagators. 

The first step in evaluating the diagram 
is to expand the integrand to second order in powers of the external 
wavevectors ${\bf p}_1$ and ${\bf p}_2$.
The expansion of the denominators has the form
\begin{eqnarray}
{1 \over \omega^2 - \epsilon^2(| {\bf k} + {\bf p} |)}
&=& 
{1 \over \omega^2 - \epsilon^2(k)}
\nonumber \\
&& \;+\; \left[
\left( {k^2 \over 2 m} + {\epsilon^2(k) \over k^2/2m} \right)
	{ p^2 + 2 {\bf p} \cdot {\bf k} \over 2m}
\;+\; {({\bf p} \cdot {\bf k})^2 \over m^2} \right]
	{1 \over (\omega^2 - \epsilon^2(k))^2}
\nonumber \\
&& \;+\; 
\left( {k^2 \over 2 m} + {\epsilon^2(k) \over k^2/2m} \right)^2
{({\bf p} \cdot {\bf k})^2 \over m^2}
	{1 \over (\omega^2 - \epsilon^2(k))^3} \,.
\end{eqnarray}
We can now use the formula (\ref{int-omega}) to evaluate the 
integral over $\omega$.  This reduces the diagram to an integral over 
${\bf k}$. We can average over the angles of ${\bf k}$
by making the substitutions $k^i k^j \to k^2 \delta^{ij}/3$ and 
$k^i \to 0$.
After simplifying the diagram, we find the terms
\begin{eqnarray}
&&
{1 \over 16 (2m)^2} X({\bf p}_1) X({\bf p}_2) \int {d^3k \over (2 \pi)^3} 
\left\{ 3 {(k^2/2m)^3 \over \epsilon^5(k)}
\right.
\nonumber \\
&& \qquad \qquad \qquad \qquad 
\left. 
\;-\; {{\bf p}_1 \cdot {\bf p}_2 \over 6 (2m)}
\left[ 35 {(k^2/2m)^6 \over \epsilon^9(k)}
	- 10 {(k^2/2m)^4 \over \epsilon^7(k)}
	+ 3 {(k^2/2m)^2 \over \epsilon^5(k)} \right]
\right\} \;.
\label{diagram-intk}
\end{eqnarray}
There are also terms proportional to 
$(p_1^2 + p_2^2)X({\bf p}_1) X({\bf p}_2)$ that we have dropped.
They correspond to terms of the form
$X \mbox{\boldmath $\nabla$}^2 X$, which first contribute to the 
density at fourth order in the gradient expansion.
Expressing the integrals over ${\bf k}$ in (\ref{diagram-intk})
in terms of the integrals $I_{m,n}$ defined in 
Appendix \ref{app:Integrals},
the expression reduces to
\begin{equation}
{3 \over 16} X({\bf p}_1) X({\bf p}_2) I_{3,5}
\;-\; {1 \over 192} \left( 35 I_{6,9} - 10 I_{4,7} + 3 I_{2,5} \right)
	{\bf p}_1 \cdot {\bf p}_2 X({\bf p}_1) X({\bf p}_2) 
 \,.
\end{equation}
After Fourier transforming to coordinate space, this becomes
\begin{equation}
{3 \over 16} I_{3,5} X^2({\bf r}) 
\;+\; {1 \over 192} \left( 35 I_{6,9} - 10 I_{4,7} + 3 I_{2,5} \right) 
(\mbox{\boldmath $\nabla$} X)^2({\bf r}) \,.
\end{equation}
>From this expression, we can now read off the coefficients $a_4$ and $a_5$
in the expansion (\ref{xi-ms:DE}) for $\langle \xi^2 \rangle_{X,Y}$.

\section{Coefficients}
\label{app:Coefficients}
\setcounter{equation}{0}

In this Appendix, we express the coefficients that appear in 
the calculation of one-loop quantum corrections to the density
in terms of the integrals $I_{m,n}$ that were defined in 
Appendix \ref{app:Integrals}.

We first list the coefficients that are used in Sections 
\ref{sec:One-loop} and \ref{sec:Self-consistent} to calculate the 
condensate and the density profile 
using the Cartesian field parameterization.
The coefficients in the expansion (\ref{xi-ms:DE}) 
for $\langle \xi^2 \rangle_{X,Y}$ are
\begin{eqnarray}
a_0 &=& {1 \over 2} I_{1,1} \,, 
\label{a0}
\\ 
a_1 &=& {1 \over 4} I_{2,3} \,, 
\\ 
a_2 &=& - {1 \over 4} I_{0,1} \,, 
\\ 
a_3 &=& {1 \over 48} \left( - 10 I_{5,7} + 13 I_{3,5} - I_{1,3} \right) \,,
\\ 
a_4 &=& {3 \over 16} I_{3,5} \,,
\\ 
a_5 &=& {1 \over 192} \left( 35 I_{6,9} - 10 I_{4,7} + 3 I_{2,5} \right) \,. 
\end{eqnarray}
The coefficients in the expansion (\ref{eta-ms:DE}) 
for $\langle \eta^2 \rangle_{X,Y}$ are
\begin{eqnarray}
b_0 &=& {1 \over 2} I_{-1,-1} \,, 
\\ 
b_1 &=& - {1 \over 4} I_{0,1} \,, 
\\ 
b_2 &=& {1 \over 4} I_{-2,-1} \,, 
\\ 
b_3 &=& {1 \over 48} \left( 2 I_{3,5} - 3 I_{1,3} - I_{-1,1} \right) \,, 
\\
b_4 &=& - {1 \over 16} I_{1,3} \,, 
\\
b_5 &=& - {5 \over 192} \left( I_{4,7} + 2 I_{2,5} + I_{0,3} \right) \,. 
\end{eqnarray}

We next list the coefficients that are required in Section \ref{sec:Polar}
to calculate the density profile using the polar field parameterization.  
The coefficients in the expansion (\ref{xixi-m}) 
for $\langle \xi^2 \rangle_{X,U,S}$ are
\begin{eqnarray}
c_0 &=& {1 \over 2} I_{1,1} \,, 
\\ 
c_1 &=& {1 \over 4} I_{3,3} \,, 
\\ 
c_2 &=& - {1 \over 4} I_{1,1} \,, 
\\ 
c_3 &=& {3 \over 16} I_{5,5} \,, 
\\ 
c_4 &=& - {1 \over 8} I_{3,3} \,, 
\\ 
c_5 &=& - {1 \over 16} I_{1,1} \,. 
\end{eqnarray}
The coefficients in the expansion (\ref{dxi-ms}) 
for $\langle ({\mbox{\boldmath $\nabla$}} \xi)^2 \rangle_{X,U,S}$ are
\begin{eqnarray}
d_0 &=& {1 \over 2} I_{2,1} \,, 
\\ 
d_1 &=& {1 \over 4} I_{3,3} \,, 
\\ 
d_2 &=& {1 \over 48} \left( - 10 I_{7,7} + 25 I_{5,5} - 9 I_{3,3} \right) \,, 
\\ 
d_3 &=& {1 \over 48} \left( 2 I_{5,5} - 7 I_{3,3} - I_{1,1} \right) \,, 
\\ 
d_4 &=& {1 \over 192} \left( 35 I_{9,9} - 90 I_{7,7} + 91 I_{5,5} \right) \,, 
\\ 
d_5 &=& {1 \over 96} \left( - 5 I_{7,7} + 14 I_{5,5} - 21 I_{3,3} \right) \,, 
\\ 
d_6 &=& {1 \over 192} \left( - 5 I_{5,5} + 6 I_{3,3}  - 13 I_{1,1} \right) \,. 
\end{eqnarray}
The coefficients in the expansion (\ref{deta-ms}) 
for $\langle ({\mbox{\boldmath $\nabla$}} \eta)^2 \rangle_{X,U,S}$ are
\begin{eqnarray}
e_0 &=& {1 \over 2} I_{0,-1} \,, 
\\ 
e_1 &=& - {1 \over 4} I_{1,1} \,, 
\\ 
e_2 &=& {1 \over 48} \left( 2 I_{5,5} - 7 I_{3,3} - I_{1,1} \right) \,, 
\\ 
e_3 &=& {1 \over 48} \left( 2 I_{3,3} - 3 I_{1,1} + 7 I_{-1,-1} \right) \,, 
\\ 
e_4 &=& {1 \over 192} \left( - 5 I_{7,7} + 6 I_{5,5} - 13 I_{3,3} \right) \,, 
\\ 
e_5 &=& {1 \over 96} \left( 3 I_{5,5} - 2 I_{3,3} - 13 I_{1,1} \right) \,, 
\\ 
e_6 &=& {1 \over 64} \left( I_{3,3} + 2 I_{1,1} + 9 I_{-1,-1} \right) \,. 
\end{eqnarray}

\newpage

\noindent
{\Large \bf Figure Caption}

\begin{description}

\item[Figure 1.]  Diagrammatic representation of the components of the 
	$2 \times 2$ propagator matrix:
	(a) the diagonal propagator for $\xi$,
	(b) the diagonal propagator for $\eta$,
	(c) the off-diagonal propagator for $\xi$ and $\eta$.

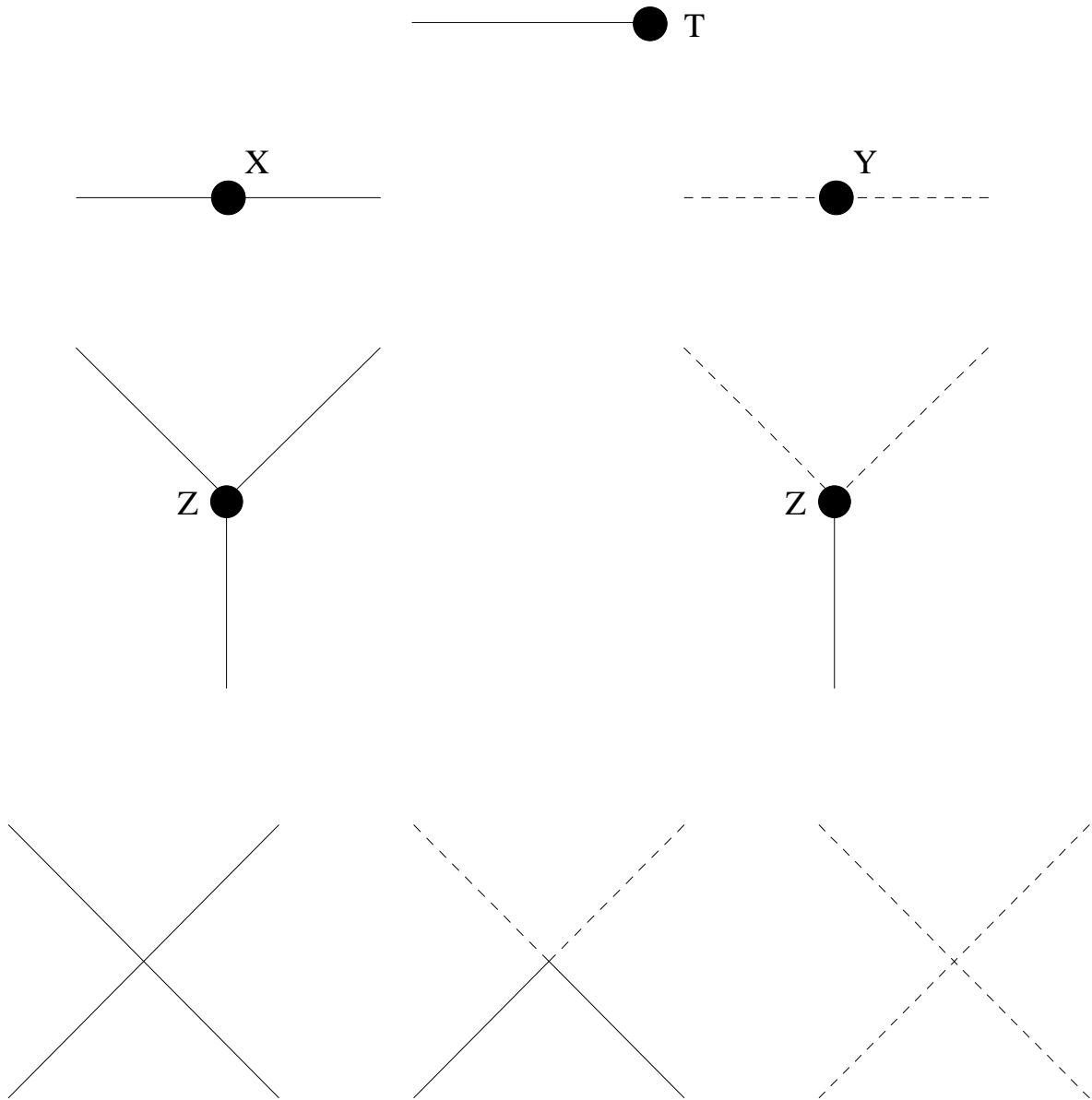
\item[Figure 2.]  Diagrammatic representation of the interaction vertices 
	associated with the sources $T$, $X$, $Y$, and $Z$ 
	and the four-point couplings.

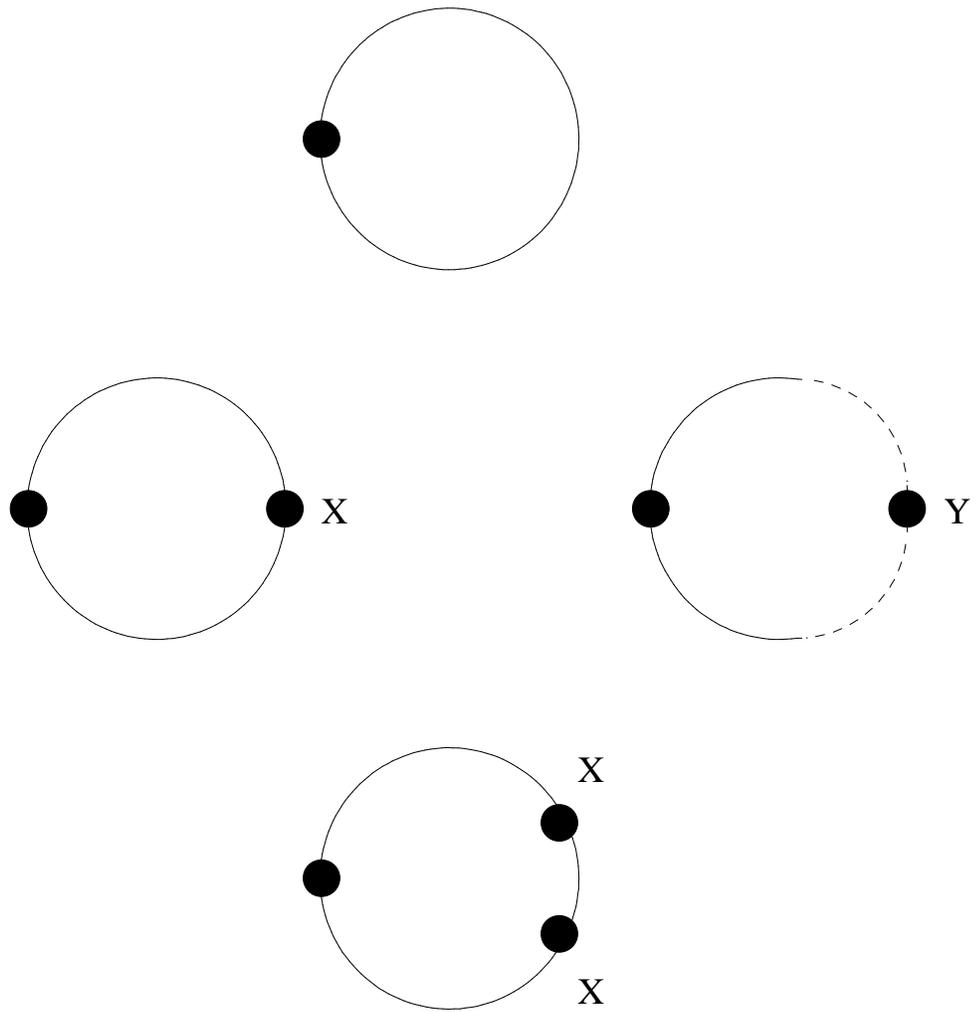
\item[Figure 3.]  One-loop Feynman diagrams that contribute to 
	$\langle \xi^2 \rangle$.

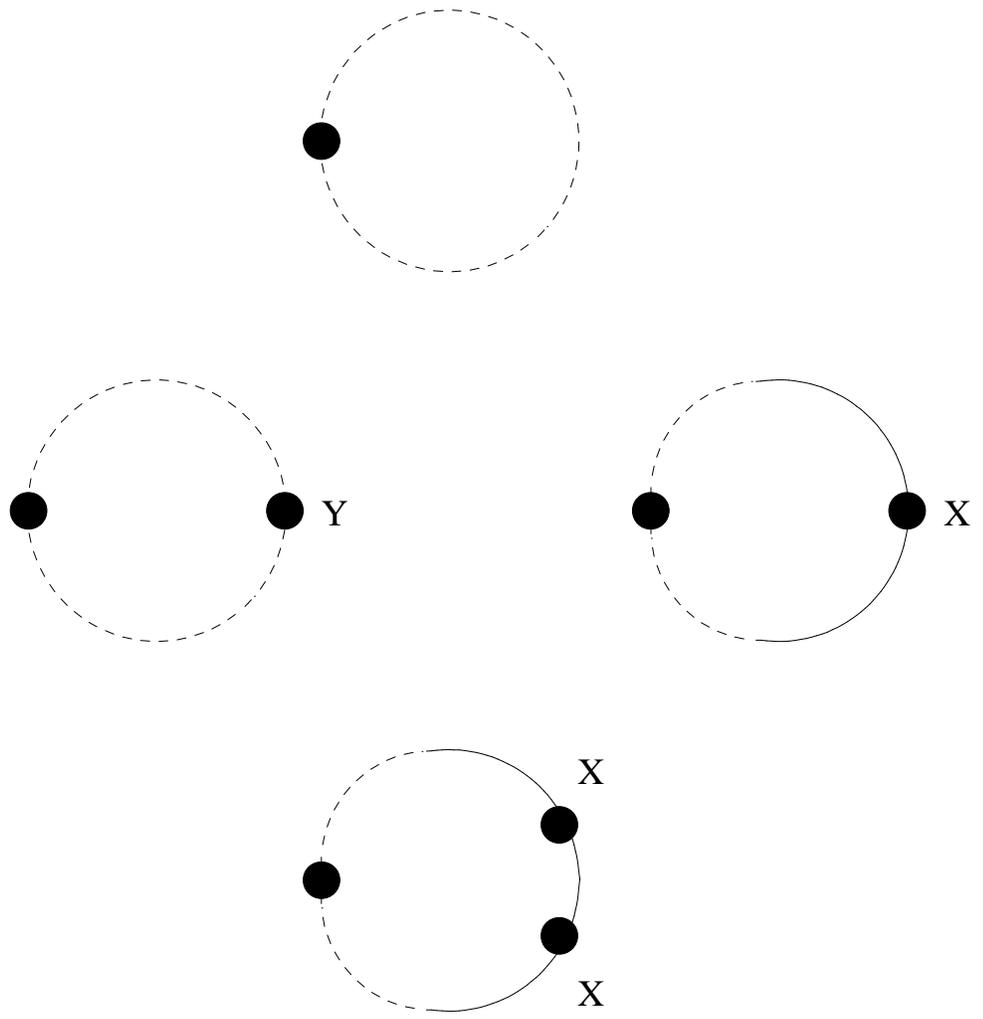
\item[Figure 4.]  One-loop Feynman diagrams that contribute to 
	$\langle \eta^2 \rangle$.

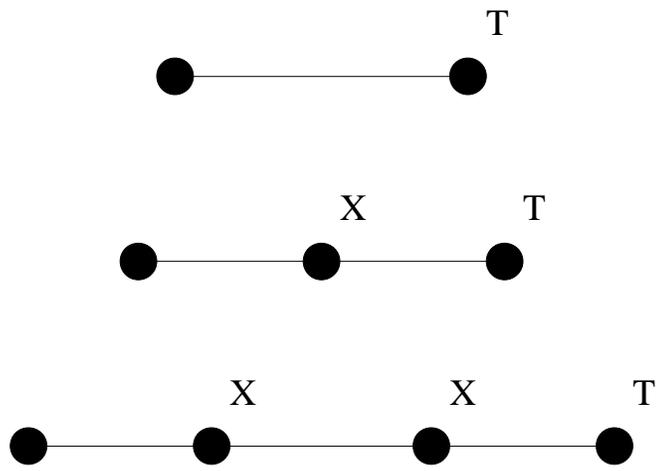
\item[Figure 5.]  Feynman diagrams contributing to 
	$\langle \xi \rangle$ that involve the source $T$.

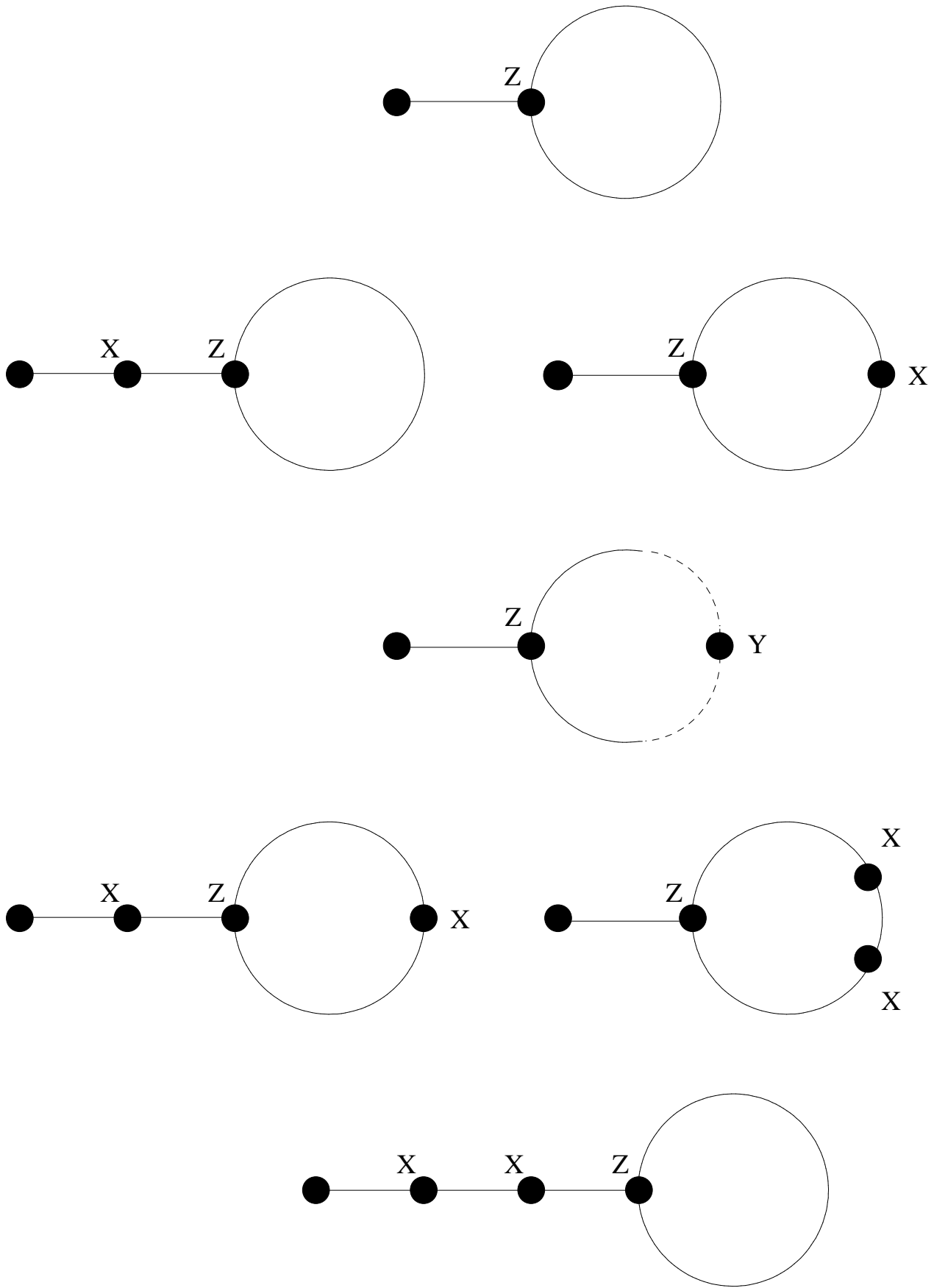
\item[Figure 6.]  One-loop Feynman diagrams contributing to 
	$\langle \xi \rangle$ that involve a pair of $\xi$ lines 
	produced by the source $Z$.

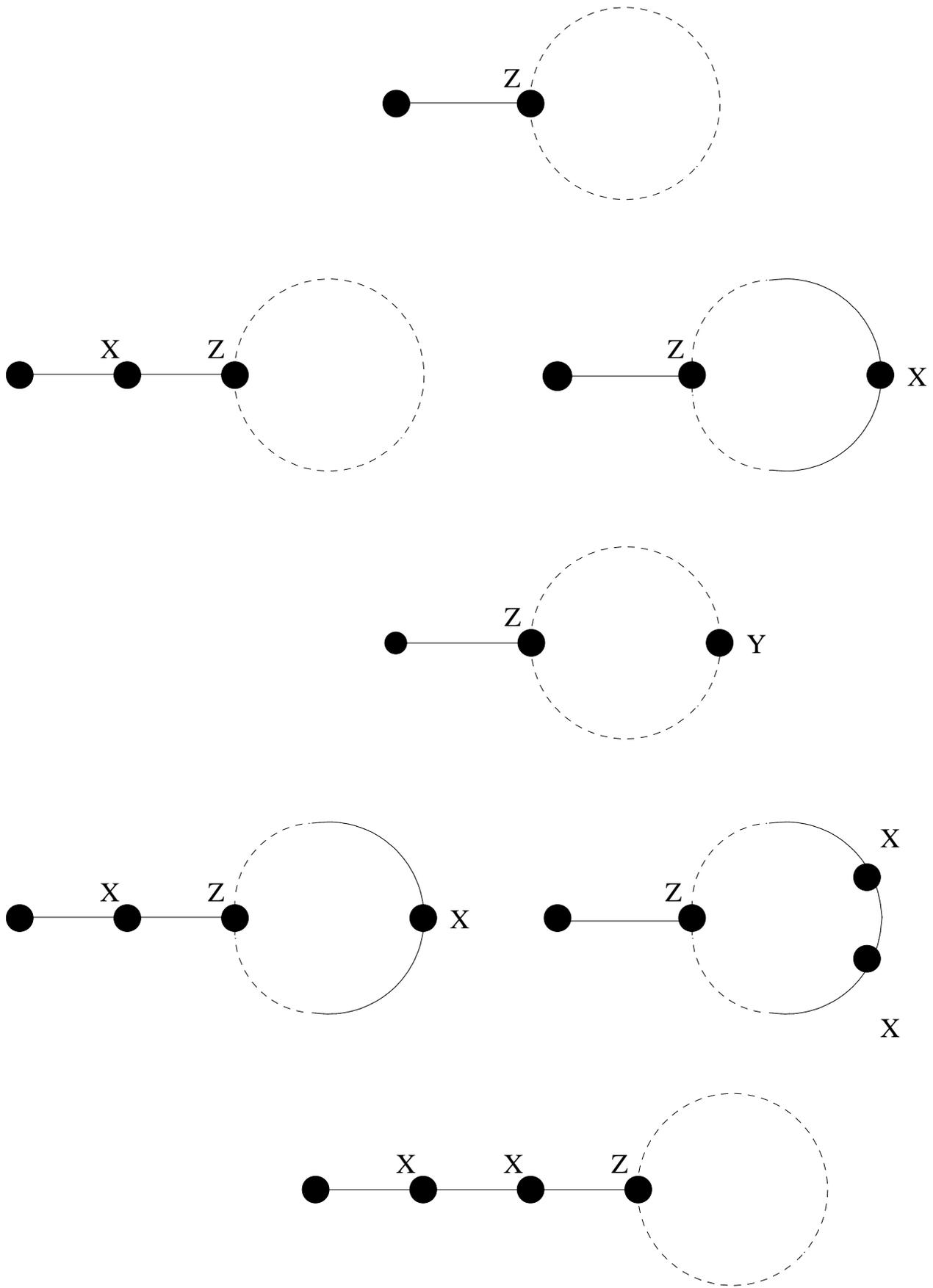
\item[Figure 7.]  One-loop Feynman diagrams contributing to 
	$\langle \xi \rangle$ that involve a pair of $\eta$ lines 
	produced by the source $Z$.

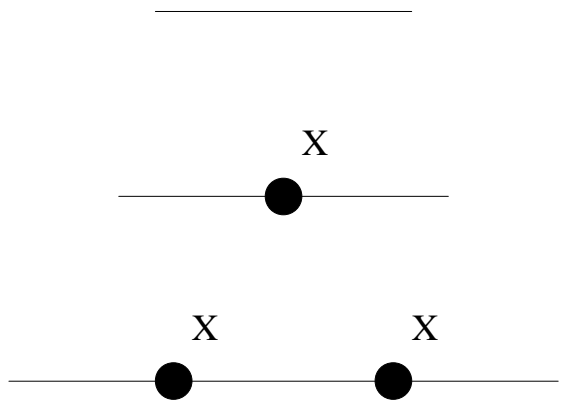
\item[Figure 8.]  Feynman diagrams that contribute to the propagator
	$D^{\xi \xi}_{X,Y}$ at zero frequency.

\end{description}

\end{document}